\documentclass[lettersize,twoside,journal]{IEEEtran}
\usepackage{graphicx}
\usepackage{amssymb}
\usepackage{amsmath,cases}
\usepackage{cite}
\usepackage{color}
\usepackage{subfigure}
\usepackage{booktabs}
\usepackage{balance}
\usepackage{acro}
\usepackage{braket}
\usepackage[mathscr]{euscript}
\usepackage{tikz}
\usetikzlibrary{quantikz2}
\usepackage{threeparttable}
\usepackage[colorlinks=true, linkcolor=black, urlcolor=blue, citecolor=black]{hyperref}
\usepackage{tabularx}
\usepackage{multirow}
\usepackage{stfloats}
\usepackage{array}
\usepackage{comment}

\newcolumntype{C}[1]{>{\centering\arraybackslash}m{#1}} 
\newcolumntype{L}{>{\raggedright\arraybackslash}X}

\usepackage{color}
\usepackage{braket}
\usepackage{bbm}
\usepackage{pifont}

\definecolor{darkgreen}{RGB}{0,128,0}

\newcommand{\M}[1]{\boldsymbol{#1}}

\newcommand{\B}[1]{\mathbf{#1}}

\DeclareMathAlphabet{\mathams}{U}{msb}{m}{n}

\newcommand{\bList}{\begin{dingautolist}{182}}
\newcommand{\eList}{\end{dingautolist}}
   
\DeclareAcronym{AI}{
short = AI,
long = artificial intelligence
}

\DeclareAcronym{CNN}{
short = CNN,
long = convolution neural network
}

\DeclareAcronym{CNOT}{
short = CNOT,
long = controlled NOT
}

\DeclareAcronym{CDLGN}{
short = CDLGN,
long = convolutional differentiable logic gate network
}

\DeclareAcronym{CV-QNN}{
short = CV-QNN,
long = continuous-variable quantum neural networks
}

\DeclareAcronym{FFN}{
short = FFN,
long = feed-forward network
}

\DeclareAcronym{FERL}{
short = FERL,
long = free energy-based reinforcement learning
}

\DeclareAcronym{FLOPS}{
short = FLOPS,
long = floating-point operations per second
}

\DeclareAcronym{HQC}{
short = HQC,
long = hybrid quantum-classical
}

\DeclareAcronym{KAN}{
short = KAN,
long = Kolmogorov Arnold network
}

\DeclareAcronym{LLM}{
short = LLM,
long = large language model
}

\DeclareAcronym{MLP}{
short = MLP,
long = multi-layer perceptrons
}

\DeclareAcronym{MHA}{
short = MHA,
long = multi-head attention
}

\DeclareAcronym{MARL}{
short = MARL,
long = multi-agent reinforcement learning
}

\DeclareAcronym{MDP}{
short = MDP,
long = Markov decision process
}

\DeclareAcronym{NISQ}{
short = NISQ,
long = noisy intermediate-scale quantum
}

\DeclareAcronym{QPU}{
short = QPU,
long = quantum processing unit
}

\DeclareAcronym{QNN}{
short = QNN,
long = quantum neural network
}

\DeclareAcronym{QEC}{
short = QEC,
long = quantum error correction
}
\DeclareAcronym{QAS}{
short = QAS,
long = quantum architecture search
}

\DeclareAcronym{QVARL}{
short = QVARL,
long = quantum variational autoencoder for reinforcement learning
}

\DeclareAcronym{QHRL}{
short = QHRL,
long = quantum hierarchical reinforcement learning
}

\DeclareAcronym{QRL}{
short = QRL,
long =  quantum reinforcement learning
}

\DeclareAcronym{QML}{
short = QML,
long = quantum machine learning
}

\DeclareAcronym{QASM}{
short = QASM,
long = quantum assembly language
}

\DeclareAcronym{QFL}{
short = QFL,
long = Quantum Federated Learning
}

\DeclareAcronym{QMARL}{
short = QMARL,
long = quantum multi-agent reinforcement learning
}

\DeclareAcronym{QAOA}{
short = QAOA,
long = quantum approximate optimization algorithm
}

\DeclareAcronym{QUBO}{
short = QUBO,
long = quadratic unconstrained binary optimization
}

\DeclareAcronym{RNN}{
short = RNN,
long = recurrent neural network
}

\DeclareAcronym{ReLU}{
short = ReLU,
long = rectified linear unit
}

\DeclareAcronym{RL}{
short = RL,
long =  reinforcement learning
}

\DeclareAcronym{SARSA}{
short = SARSA,
long = state-action-reward-state-action
}

\DeclareAcronym{SDK}{
short = SDK,
long = software development kit
}

\DeclareAcronym{SAGIN}{
short = SAGIN,
long = Space-Air-Ground Integrated Network
}

\DeclareAcronym{TD}{
short = TD,
long = temporal-difference
}

\DeclareAcronym{UAV}{
short = UAV,
long = Unmanned Aerial Vehicle
}

\DeclareAcronym{VQC}{
short = VQC,
long = variational quantum circuit
}

\DeclareAcronym{VQE}{
short = VQE,
long = variational quantum eigensolver
}

\begin{document}

\title{
Quantum Reinforcement Learning: Recent Advances and Future Directions
}

\author{Jawaher~Kaldari,
Shehbaz~Tariq,
Saif~Al-Kuwari,~\IEEEmembership{Senior Member,~IEEE},
Samuel~Yen-Chi~Chen, 
Symeon~Chatzinotas,~\IEEEmembership{Fellow,~IEEE},
Hyundong~Shin,~\IEEEmembership{Fellow,~IEEE}

\thanks{J.~Kaldari, and S.~Al-Kuwari are with the Qatar Center for Quantum Computing, College of Science and Engineering, Hamad Bin Khalifa University, Doha, Qatar (e-mail: jaka51804@hbku.edu.qa; smalkuwari@hbku.edu.qa).}
\thanks{S.~Tariq and S.~Chatzinotas are with the Interdisciplinary Centre for Security, Reliability and Trust (SnT), University of Luxembourg, 1855 Luxembourg City, Luxembourg (e-mail: shehbaz.tariq@uni.lu; symeon.chatzinotas@uni.lu).}
\thanks{S. Y.-C. Chen is with the Wells Fargo, New York, NY, USA (e-mail: ycchen1989@ieee.org)}
\thanks{H.~Shin 
is with the Department of Electronics and Information Convergence Engineering,
Kyung Hee University,
1732 Deogyeong-daero, Giheung-gu,
Yongin-si, Gyeonggi-do 17104,
Republic of Korea (e-mail: hshin@khu.ac.kr).}
\thanks{J.~Kaldari, and S.~Tariq contributed equally to this paper.
}

\thanks{The views expressed in this article are those of the authors and do not represent the views of Wells Fargo. This article is for informational purposes only. Nothing contained in this article should be construed as investment advice. Wells Fargo makes no express or implied warranties and expressly disclaims all legal, tax, and accounting implications related to this article.}
}
\maketitle

\begin{abstract}
As quantum machine learning continues to evolve, reinforcement learning stands out as a particularly promising yet underexplored frontier. In this survey, we investigate the recent advances in \ac{QRL} to assess its potential in various applications. While \ac{QRL} has generally received less attention than other quantum machine learning approaches, recent research reveals its distinct advantages and transversal applicability in both quantum and classical domains. We present a comprehensive analysis of the \ac{QRL} framework, including its algorithms, architectures, and supporting \acp{SDK}, as well as its applications in diverse fields. Additionally, we discuss the challenges and opportunities that QRL can unfold, highlighting promising use cases that may drive innovation in quantum-inspired reinforcement learning and catalyze its adoption in various interdisciplinary contexts.

\end{abstract}

\begin{IEEEkeywords}
quantum computing, reinforcement learning, quantum machine learning, variational quantum circuits, quantum optimization
\end{IEEEkeywords}



\acresetall		

\section{Introduction}
\label{sec:1}
\IEEEPARstart{T}{he} current generation of \ac{NISQ} devices, consisting of hundreds of qubits, is expected to enable computations beyond the reach of today’s classical supercomputers \cite{Pre:18:Quantum}. Different approaches are being pursued to develop these \ac{NISQ} devices, including superconducting systems\cite{NRK:18:Science}, trapped ion systems \cite{ZPHKBKGGM:17:Nat}, quantum dots \cite{LD:98:PRA}, cold atomic arrangements \cite{BSKLOPCZEG:17:Nat}, and photonic computing platforms \cite{RZBB:94:PRL}. These devices are expected to achieve quantum supremacy in specific applications, tackling computations that would be infeasible for classical computers, thereby revealing new opportunities in scientific research and industrial applications \cite{NKKUB:24:arXiv, Ter:18:NPh, MLARVBM:22:Nat, AABBBBB:19:Nat, HMCP:2024:Comm_Phys, AHHG:23:PRX_Quantum}. However, significant challenges remain, mainly due to the inherent noise and decoherence in quantum gates, which limit the robustness and fidelity of quantum computations that enable the execution of more complex algorithms compared to state-of-the-art classical systems \cite{RK:21:ACM_CSUR}. 

\Acp{VQC} are widely employed to demonstrate a near-term quantum advantage in the \ac{NISQ} era. These parameterized quantum circuits are well-suited for current quantum technology due to their adaptability to noisy hardware and support for hybrid quantum-classical workflows \cite{CABBEF:21:NRP}. Remarkably, noise within \acp{VQC} can enhance exploration during optimization, a critical advantage that can be harnessed by \ac{QRL} \cite{KB:23:PRA, JGMBD:21:NeurIPS}. By leveraging noise constructively, \ac{QRL}, supported by \acp{VQC}, enables efficient learning in complex environments where classical reinforcement learning struggles \cite{HDCHG:22:arXiv, KTANLN:2023:arXiv}. 

Recent advances highlight the potential of \ac{VQC}-based \ac{QRL} to achieve quantum advantage even under noisy \ac{NISQ} conditions. Through parameter-efficient quantum policies, quantum parallelism, and robust optimization, \ac{QRL} offers faster convergence and enhanced performance in high-dimensional or noisy environments, making it particularly well-suited for resource-constrained and dynamic systems\cite{SMBMD:23:EPJ_QT}. In fact, some types of noise can improve algorithmic effectiveness, promoting exploration across large action spaces \cite{JTNBD:21:PRX_Quantum}. Recent experimental findings further confirm quantum speed-ups in learning, validating \ac{QRL} for complex decision-making tasks \cite{SAHSSDFHHEWBW:21:Nature}. Beyond decision-making, quantum-inspired \ac{RL} techniques are advancing diverse quantum applications, including quantum architecture search \cite{kuo2021quantum}, quantum sensing \cite{SFB:20:NJP}, and quantum control \cite{NBSN:19:npj_QI}. These developments underscore the versatility of \ac{RL} in enhancing quantum technologies.


RL has been extensively studied for decades in the classical domain, leading to a wide range of theoretical and practical advancements. In contrast, its counterpart in the quantum domain is a much more recent development. Despite growing interest in QRL, the number of comprehensive surveys in the literature remains limited. Table~\ref{tab:SurveyComparison} summarizes several existing surveys and compares them to ours.

\begin{table*}[ht]
    \centering
    \caption{Previous Survey Comparison and contributions}
    \label{tab:SurveyComparison}
    \begin{tabular}{ccccccc}
    \toprule
        \textbf{Ref.} & \textbf{Tutorials}  & \textbf{Architectures} & \textbf{QRL Applications} & \textbf{RL Applications} & \textbf{Benchmarking} \\
        \midrule
        \cite{alomari2025survey} & $\times$  & $\checkmark$ & $\checkmark$ & $\O$ & $\times$\\

        \cite{meyer2022survey} & $\times$  & $\checkmark$ & $\checkmark$ & $\times$ & $\times$\\
        \cite{park2024trends} & $\times$  & $\O$ & $\checkmark$ & $\times$ &$\times$ \\
       
        \cite{yu2023quantum} & $\times$  & $\O$ & $\O$ & $\times$ & $\times$ \\   

        \cite{chen2024introduction} & $\times$  & $\checkmark$ & $\O$ & $\times$ & $\times$ \\

        \cite{chen2026quantum}  & $\times$  & $\checkmark$ & $\O$ & $\times$ & $\times$ \\

        This Work & $\checkmark$  & $\checkmark$ & $\checkmark$ & $\checkmark$ & $\checkmark$ \\

        \bottomrule
    \end{tabular}
    \\[1ex]
    \textbf{Legend:} $\checkmark$= Covered in detail, $\times$ = Not mentioned, $\O$ = Briefly mentioned.
    \label{tab:placeholder}
\end{table*}

The rest of this survey is organized as follows. Section \ref{sec:Pre} reviews a few fundamental concepts to establish the theoretical basis for \ac{QRL}. Section \ref{sec:QRL} introduces the \ac{QRL} framework, detailing its integration with \acp{VQC} and their role in achieving quantum advantage. Section \ref{sec:Arch} describes the main QRL architectures, while Section \ref{sec:Algorithms} examines the QRL algorithms, as well as short tutorials. Section \ref{sec:benchmark} discusses benchmarking issues and recent advances in this area. Sections \ref{sec:classicalRLappl} and \ref{sec:QRLApp} present applications of classical RL to quantum systems and applications of QRL itself, respectively. Section \ref{sec:future} highlights key challenges and outlines promising future directions. Finally, Section \ref{sec:conclusion} concludes the survey.


\section{Preliminaries}
\label{sec:Pre}

\subsection{Reinforcement Learning}
\Ac{RL} is a computational approach in which an agent learns to make sequential decisions by interacting with an environment to maximize cumulative rewards, as shown in Fig.~\ref{fig:1}. This process is commonly modeled as a \ac{MDP} \cite{SB:18:MIT}, characterized by:
\begin{figure}
    \centering
    \includegraphics[width=0.98\linewidth]{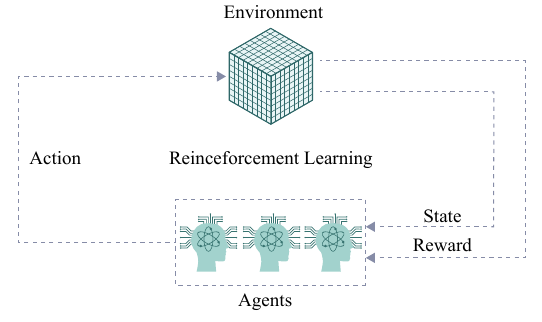}
    \caption{Reinforcement Learning cycle where the agents recursively interact with their environment and learn by associating rewards with their actions.}
    \label{fig:1}
\end{figure}
\begin{itemize}
    \item A set of states $ \mathcal{S} $, representing the environment's possible conditions.
    \item A set of actions $ \mathcal{A} $, defining the choices available to the agent.
    \item A transition function $ \mathcal{P}: \mathcal{S} \times \mathcal{A} \times \mathcal{S} \rightarrow [0,1] $, where $ \mathcal{P}(s'|s,a) $ denotes the probability of transitioning to state $ s' $ from state $ s $ after taking action $ a $.
    \item A reward function $ \mathcal{R}: \mathcal{S} \times \mathcal{A} \times \mathcal{S} \rightarrow \mathbb{R} $, providing feedback on the agent's actions to guide its behavior.
\end{itemize}

At each discrete time step $t$, the agent observes the current state $s_t \in \mathcal{S}$ and selects an action $a_t \in \mathcal{A}$ based on a policy $\pi$, which can be deterministic or stochastic. The environment then transitions to a new state $s_{t+1}$ according to the transition function $ \mathcal{P} $, and the agent receives an immediate reward $r_t = \mathcal{R}(s_t, a_t, s_{t+1})$. This immediate reward provides direct feedback on the outcome of the agent's action at that specific time step.
However, the agent's goal is not just to maximize immediate rewards, but to learn behaviors that lead to high cumulative rewards over time. This is captured by the expected cumulative reward, often called the return, and is defined as: 
\begin{align}
    R_t = \sum_{\tau=0}^\infty \gamma^\tau r_{t+\tau},
\end{align}
where $\gamma$ is a discount factor between $0$ and $1$ that determines the importance of future rewards. If $\gamma=0$, the expected reward reduces to $R_t=r_t$, which means that the agent will only care about the immediate rewards and ignore future ones. This may cause the agent to favor short-term rewards and ignore strategies that lead to better outcomes in the long run. On the other hand, if $\gamma$ approaches 1, the agent will give almost equal importance to future and immediate rewards, which encourages long-term planning. However, this can result in unstable learning or divergence in infinite-horizon tasks. Therefore, the choice of $\gamma$ plays a critical role in balancing short-term and long-term objectives and is essential for learning effective policies in reinforcement learning.

The agent's goal is to find a policy that maximizes the expected cumulative rewards. The policy is formally defined as a function that maps each state to an action. In the simplest case, a deterministic policy maps each state to a specific action $a=\pi(s)$, which limits the agent's ability to explore alternative actions that might lead to higher long-term rewards. In contrast, a stochastic policy maps each state to a probability distribution over actions.

\begin{equation}
    \pi(a \mid s) = P[\mathcal{A} _t = a \mid \mathcal{S} _t = s]
\end{equation}

If an agent follows a policy $\pi$, then at time step $t$, it selects action 
$a \in \mathcal{A}$ given that $\mathcal{S}_t=s$ at time $t$. Since $\pi(a \mid s)$ defines a valid probability distribution, it must satisfy the normalization condition:

\begin{equation}
    \sum_{a \in \mathcal{A}} \pi(a \mid s) = 1 \; \forall s \in \mathcal{S}
\end{equation}

Stochastic policies allow the agent to explore multiple actions by occasionally selecting the ones that are not currently considered best, but that might lead to better long-term rewards, rather than always committing to a single deterministic choice.

Since the agent's goal is to learn a policy that maximizes the expected cumulative rewards, the agent must be able to evaluate how \emph{good} each state and action is in the long term. This is captured through the value functions, which estimate the expected return associated with states or state-action pairs under a given policy. There are mainly two types of value functions:

\begin{itemize}
    \item \textit{State-Value Function:} The state-value function tells the agent how good it is to be in a specific state. It can be formally defined as the expected return when starting in state $s$ and following policy $\pi$:
    \begin{align}
        v_\pi(s) &= \mathbb{E}_\pi \left[ r_t \mid S_t = s \right]  \nonumber \\ 
         &= \mathbb{E}_\pi \left[ \sum_{\tau=0}^{\infty} \gamma^\tau R_{t+\tau+1} \mid S_t = s \right].
    \end{align}

    \item \textit{Action-Value Function}: The action-value (also known as the Q-function) function tells the agent how good it is to perform a specific action in a specific state. It is formally defined as the expected return when starting in state $s$, taking action $a$, and following policy $\pi$ afterwards: 
    \begin{align}
    q_\pi(s, a) &= \mathbb{E}_\pi \left[ R_t \mid S_t = s, A_t = a \right] \nonumber \\
    &= \mathbb{E}_\pi \left[ \sum_{\tau=0}^{\infty} \gamma^k r_{t+\tau+1} \mid S_t = s, A_t = a \right].
    \end{align}
    If an agent knows the $q_\pi(s, a)$ values for every action $a$ in a given state $s$, it can easily find the best action, which corresponds to the one with the highest Q-value. 

\end{itemize}

The ultimate goal in RL is to find the optimal policy, denoted $\pi^*$. An optimal policy is defined as a policy that is better than or equal to all other policies. This condition is satisfied if the expected return under $\pi^*$ is greater than or equal to the expected return under any other policy $\pi$, for all possible states. This can be formally defined using the state-value function:

\begin{equation}
    v_{\pi^*}(s) \geq v_\pi(s) \quad \text{for all } s \in \mathcal{S}
\end{equation}

Therefore, the optimal policy leads to the optimal state value function $v_{\pi^*}(s)$, defined as $v_{\pi^*}(s) = \max_{\pi} v_\pi(s)$, and the optimal action-value function $q^*_\pi(s, a)$, defined as $q^*_\pi(s, a) = \max_{\pi} q_\pi(s, a)$. 

A powerful property of these optimal value functions is that they satisfy the Bellman optimality equations (these are key recursive equations often used in RL to find the maximum possible future reward an agent can achieve from a given state). The Bellman optimality equation for the optimal action-value function is:
\begin{equation}
    q^*(s, a) = \mathbb{E} \left[ r_{t+1} + \gamma \max_{a'} q_*(s', a') \right]
\end{equation}
where $s'$ denotes the next state resulting from taking action $a$ in state $s$, and $a'$ denotes the possible actions available in state $s'$. 

Many RL algorithms are built upon Bellman's equation, including Q-learning. In Q-learning, the agent's goal is to find the optimal action-value function $q^*_\pi(s, a)$ that satisfies the Bellman optimality. To approximate this function, Q-learning maintains a table of Q-values, a lookup table initialized with all zeros, which stores the estimated Q-values for each state-action pair. As the agent interacts with the environment, it iteratively updates these values using the temporal difference learning rule:
\begin{equation}
    Q(s, a) \leftarrow Q(s, a) + \alpha[r + \gamma \max_{a'}Q(s', a') - Q(s, a)]
\end{equation}
where $\alpha$ is the learning rate. The final learned policy is simple and deterministic: in any given state, the agent selects the action that maximizes the Q-value from the table. Over time, this iterative process leads to convergence towards optimal Q-values. This tabular approach makes classical Q-learning highly effective for problems with small, discrete state and action spaces. However, as the environment's state or action space becomes larger or continuous, maintaining and updating the Q-table becomes infeasible. To address this limitation, Deep Q-learning (DQN) replaces the Q-table with a neural network that approximates the Q-function \cite{mnih2015human}. Despite this increased complexity, deep Q-networks follow the same principle: select the action with the highest predicted Q-value. 

\begin{figure*}
    \centering
    \includegraphics[width=1\linewidth]{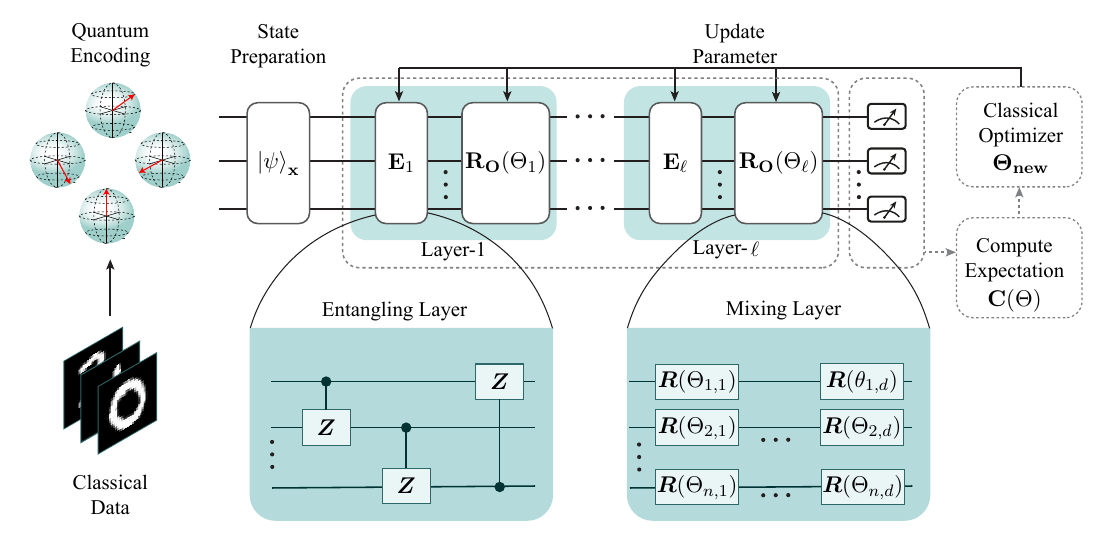}
    \caption{General framework of a \ac{VQC}, illustrating parameterized unitary operations for applications in optimization, learning, and quantum-enhanced tasks.}
    \label{fig:VQC}
\end{figure*}
\subsection{Variational Quantum Circuit}
In quantum computing, a sequence of unitary operators forms a quantum circuit. Introducing trainable parameters into these circuits results in \acp{VQC}, enabling the circuits to learn various tasks such as optimization and approximation \cite{CAB:21:NRP}. \acp{VQC} have found applications in various areas, including \ac{QRL} \cite{chen2020variational}, \ac{VQE} \cite{CSAC:22:NPJ_QI}, Quantum Generative Models \cite{TSDZLYWHWW:23:IEEE_J_PAMI}, and \acp{QNN} \cite{HCTHKCG:19:Nat}, as shown in Fig.~\ref{fig:VQC}. A key element of \acp{VQC} is the ansatz—the specific structure of parameterized unitary operators. The ansatz structure may vary by task, but typically includes parameterized unitary operators in the form:
\begin{align}
    \B{R_O}(\Theta) = \exp{\left(-i \B{O} \frac{\Theta}{2}\right)},
\end{align}
where $ \B{O} \in \{\B{X}, \B{Y}, \B{Z}\} $ represents Pauli matrices.

To process classical data with \acp{VQC}, it is essential to encode the data into quantum states. Common encoding methods include basis encoding, amplitude encoding, and angle encoding \cite{SP:21:ML_QC}. 

\begin{itemize}
    \item \textbf{Basis encoding:} Each classical bit string is mapped directly to a computational basis state of the qubits.
    \item \textbf{Amplitude encoding:} A normalized classical vector is encoded into the amplitudes of a quantum state.
    \item \textbf{Angle encoding:} Classical data values are mapped to the parameters of quantum rotation gates (e.g., $R_x$, $R_y$, $R_z$).
\end{itemize}

A comparative summary of these encoding methods is presented in Table \ref{tab:Encoding}.

\begin{table*}[t!]
\centering
\caption{Summary of Quantum Data Encoding Methods}
\label{tab:Encoding}
\setlength{\tabcolsep}{6pt}
\renewcommand{\arraystretch}{1.25}
\begin{tabularx}{\textwidth}{C{3.0cm} L L}
\toprule
\textbf{Encoding Method} & 
\multicolumn{1}{c}{\textbf{Key Advantages}} & 
\multicolumn{1}{c}{\textbf{Key Disadvantages}} \\
\midrule

\multirow{5}{*}{\textbf{Basis Encoding}}
& Direct mapping of binary data to computational basis states, enabling straightforward implementation. 
& Linear growth of qubit count with data dimension, leading to high resource demand. \\

& Requires only state initialization without complex gates. 
& Limited expressiveness—fails to capture correlations between data bits. \\

& Offers transparent data interpretability and debugging simplicity.
& Significant initialization overhead for large datasets. \\

\addlinespace
\midrule
\addlinespace

\multirow{5}{*}{\textbf{Angle Encoding}}
& Efficient in qubit usage by mapping classical features to rotation angles. 
& Increased circuit depth due to sequential rotation gates. \\

& Naturally supports continuous data through tunable gate parameters. 
& Sensitive to calibration errors and gate noise affecting encoded precision. \\

& Easily implemented on \ac{NISQ} hardware with standard single-qubit rotations.
& Overlapping state representations may reduce class separability for large feature sets. \\

\addlinespace
\midrule
\addlinespace

\multirow{5}{*}{\textbf{Amplitude Encoding}}
& Most qubit-efficient method—encodes $2^n$ data values in $n$ qubits. 
& Complex state preparation requiring multi-controlled rotations or quantum memories. \\

& Preserves continuous relationships between data values via amplitude ratios. 
& Normalization step may distort relative feature scales or lose information. \\

& Compatible with advanced techniques such as quantum singular-value transformation and data re-uploading.
& Deep entangling circuits increase noise accumulation and error-correction overhead. \\

\bottomrule
\end{tabularx}
\end{table*}

Once classical data is encoded into quantum states, the \ac{VQC} is trained for specific tasks by adjusting the parameters of the quantum gates to minimize a predefined cost function (the latter measures the discrepancy between the circuit’s output and the desired result). The training process for a \ac{VQC} involves the following steps:

\begin{enumerate}
    \item \textit{Initialization:} Initialize the quantum gate parameters, represented as $\Theta = \{\Theta_1, \Theta_2, \ldots, \Theta_n\}$, either randomly or based on prior knowledge.

    \item \textit{Forward Pass:} Encode classical data $x$ into quantum states, then apply the parameterized unitary transformations $\B{U}(\Theta)$ to evolve the initial state $\ket{0}$ into the output state:
    \begin{align}
        \ket{\M{\psi}_{\text{out}}} = \B{U}(\Theta) \ket{\M{\psi}_{\text{in}}}.
    \end{align}

    \item \textit{Measurement:} Measure the output state $\ket{\M{\psi}_{\text{out}}}$ in a chosen basis to extract classical information, yielding measurement outcomes $m$ (e.g., probabilities or expectation values).

    \item \textit{Cost Evaluation:} Calculate the cost function $C(\Theta)$ using the measured outputs $m$ and target values $y$. For example, the cost function may be given by:
    \begin{align}
        C(\Theta) = \braket{m | \hat{O} | m} - y,
    \end{align}
    where $\hat{O}$ is an observable operator corresponding to the measurement, and $y$ is the target value.

   \item \textit{Parameter Update:} Adjust the parameters $\Theta$ to minimize $C(\Theta)$. This can be achieved using optimization algorithms, such as gradient descent or gradient-free methods. For gradient-based approaches, the update rule:
    \begin{align}
        \Theta_{new} \leftarrow \Theta - \eta \nabla_\Theta C(\Theta),
    \end{align}
    where $\eta$ is the learning rate. In gradient-free methods, parameters are updated based on alternative strategies, such as evolutionary algorithms or sampling techniques.

    \item \textit{Iteration:} Repeat the forward pass, measurement, cost evaluation, and parameter update steps until the cost function $C(\Theta)$ converges to an acceptable minimum.
\end{enumerate}

\section{Quantum Reinforcement Learning}
\label{sec:QRL}

\Ac{QRL} extends classical \ac{RL} by integrating quantum computing, allowing agents to interact with quantum environments to maximize cumulative rewards and improve learning performance and efficiency compared to classical reinforcement learning approaches. The authors in \cite{JTNBD:21:PRX_Quantum} demonstrate that a hybrid quantum-classical approach, leveraging quantum-enhanced sampling and energy-based models, achieves superior learning performance over classical deep \ac{RL}, especially in large action-space environments. Similarly, the authors in \cite{SAHSSDFHHEWBW:21:Nature} show a quantum speed-up in learning times through quantum communication channels, reducing the epochs needed to reach optimal performance. This framework establishes quantum states, actions, transition operators, and reward operators within Hilbert spaces, highlighting a systematic quantum advantage in \ac{RL}. 
\begin{figure}
    \centering
    \includegraphics[width=1\linewidth]{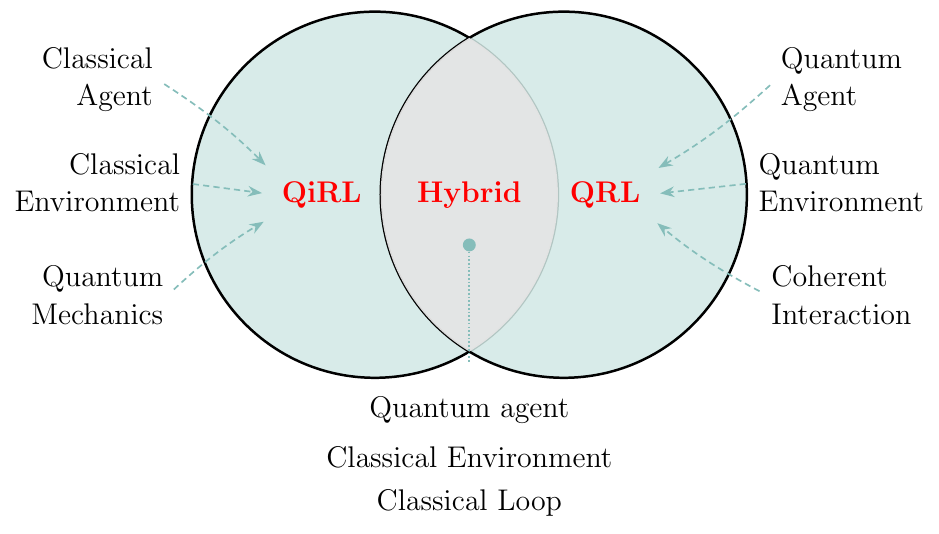}
    \caption{Taxonomy of Quantum Reinforcement Learning approaches}
    \label{fig:taxonomy}
\end{figure}
\subsection{Taxonomy}
In practice, QRL approaches can be categorized into three categories, as illustrated in Figure \ref{fig:taxonomy}:
\begin{enumerate}
    \item Quantum-Inspired RL (QiRL): Entirely classical algorithms that borrow principles from quantum mechanics to enhance exploration or optimization. This is discussed in detail below. 
    \item Hybrid Quantum-Classical: The RL loop remains classical, but certain components, such as the policy or value function, are replaced with parameterized quantum circuits. This is the most common approach in the current literature and is covered in detail in this survey. 
    \item Fully Quantum RL: All components of the pipeline are quantized. Both the agent and the environment are treated as quantum systems that interact coherently, allowing superpositions of trajectories and the use of algorithms such as Grover search. These methods are mainly theoretical (at this point in time) and generally require fault-tolerant to fully observe quantum advantage. 
\end{enumerate}

Quantum-Inspired Reinforcement Learning (QiRL) differs significantly from standard QRL. In QRL, the algorithms are designed to run on quantum hardware, leveraging quantum circuits to represent policies or value functions. In contrast, QiRL takes inspiration from quantum mechanics but develops algorithms that are entirely classical and are executed on classical computers. Several quantum phenomena have been borrowed from classical RL, enabling improved exploration, optimization, and decision-making strategies without the need for quantum devices. Examples of quantum mechanical phenomena adopted by QiRL include:

\begin{itemize}
    \item \textit{Amplitude Amplification:} Amplitude amplification, used in Grover’s algorithm, increases the amplitude of the quantum state corresponding to the correct solution, thus increasing the probability of measuring that solution. In QiRL, amplitude-inspired methods adapt this idea to boost the selection probability of high-reward actions \cite{5669349}. 
    \item \textit{Collapse phenomenon:} In quantum mechanics, measurement causes a quantum state in superposition to collapse into one of its basis states, with the probability of each outcome given by the square of its amplitude. In QiRL, this idea is adapted for the selection of probabilistic actions, where the agent chooses actions based on a learned probability distribution, encouraging exploration rather than always selecting the highest reward action \cite{5669349}.
    \item \textit{Quantum annealing:} Quantum annealing is a quantum optimization method designed to find the global minimum of a given cost function by exploiting quantum mechanics, especially quantum tunneling. In QiRL, annealing-inspired schedules are used to escape local optima in a large search space \cite{wang2024quantum}.
    \item \textit{Quantum walks:} Quantum walks inspire RL exploration strategies in which the agent searches the state space in a way that mimics quantum superposition and interference, allowing faster or more efficient coverage of possible states than purely random exploration \cite{paparo2014quantum}. 
\end{itemize}

Fully QRL methods have been proposed in the literature, but they remain largely theoretical. In \cite{dunjko2015framework}, the authors proposed a general framework for fully quantum reinforcement learning in which both the agent and the environment are modeled as quantum systems. The agent and environment each have internal quantum registers and exchange information through completely positive trace-preserving (or unitary) maps, allowing queries of the environment in superposition over action sequences so the agent can learn in parallel. To enable such superposed queries, the environment must be oracularized (i.e., it should behave as a quantum oracle that coherently encodes rewards). Several papers extended and generalized these ideas, exploring oracular access and conditions for provable speed-ups \cite{dunjko2017advances,hamann2021quantum,wang2021quantum}.

\subsection{Definition}
To formally define the \ac{QRL} framework, the key distinction from classical reinforcement learning lies in the quantum representation of both the state and the policy. In \ac{QRL}, the environment’s state at each time step $t$ is encoded as a quantum state ${\ket{\psi_t}}$ in the Hilbert space $\mathcal{H}_\mathcal{S}$.  The agent’s policy $\B{U}_\Theta$, a unitary transformation parameterized by $\Theta$, maps the observed states to actions by measuring:

\begin{align}
    \B{U}_\Theta \ket{\M{\psi}_t}
\end{align}

After performing the action, the agent receives a scalar reward $r_t$ and the environment transitions to a new quantum state $\ket{\psi_{t+1}}$. The objective is to optimize the policy parameters $\Theta$ to maximize the expected cumulative reward: 

\begin{align}
    R_t = \sum_{\tau=0}^{\infty} \gamma^\tau r_{t+\tau}
\end{align}
where $0 < \gamma \leq 1$ is the discount factor. The agent maximizes expected cumulative rewards across the trajectory:
\begin{align}
    \mathbb{E}\left[ R_t \right] = \mathbb{E} \left[ \sum_{\tau=0}^{\infty} \gamma^\tau r_{t+\tau} \right]
\end{align}
Optimizing the parameters $\Theta$ in $\B{U}_\Theta$ to achieve this yields:
\begin{align}
    \Theta \leftarrow \Theta + \eta \nabla_\Theta \mathbb{E}\left[ R_t \right]
    \label{eq:update}
\end{align}

\subsection{Software Frameworks}
\Acp{SDK}s are essential for advancing research in \ac{QRL}, offering foundational tools, libraries and environments that support the development, testing and deployment of quantum algorithms. These frameworks provide critical features such as differentiable programming, enabling the optimization of \ac{QRL} models by allowing gradients to flow through quantum circuits and facilitating hybrid quantum-classical workflows\cite{JTKWLGMBNC:2024:arXiv, BISGAAAANAA:2018:arXiv, CJAABHRSG:22:ACM_TQC}. High performance simulation capabilities in many \acp{SDK} further enhance research by allowing experimentation with complex quantum algorithms in controlled environments, enabling iterative development and testing before deployment on actual quantum hardware \cite{A:2024:arXiv}. As shown in Table~\ref{tab:SDK}, frameworks such as Qiskit, PennyLane, and TensorFlow Quantum are particularly valuable in the \ac{QRL} ecosystem. These \acp{SDK} offer high-level abstractions and integrate seamlessly with classical machine learning libraries, making it easier to build \ac{QRL} models. For example, Qiskit and PennyLane support GPU acceleration and integrate with popular ML libraries, while CUDA Quantum and TorchQuantum leverage NVIDIA GPUs for enhanced simulation capabilities. The unique features of each framework, including hardware backends, machine learning integration, and availability of \ac{QRL}-specific tools, make them crucial for researchers aiming to build efficient and scalable quantum-enhanced reinforcement learning models.

\section{QRL Architectures}\label{sec:Arch}
This section presents an overview of four advanced architectures in quantum reinforcement learning: Quantum Multi-Agent RL (QMARL), Free-Energy RL (FERL), Quantum Variational Autoencoder RL (QVARL), and Quantum Hierarchical RL (QHRL). For each architecture, we outline the fundamental idea and highlight representative papers that illustrate its development and applications.

\subsubsection{Quantum multi-agent reinforcement learning}
The Quantum multi-agent reinforcement learning (QMARL) framework extends classical \ac{RL} to settings where multiple agents interact within a shared quantum environment \cite{yun2023quantum}. Let $\ket{\M{\psi}_{i, t}} \in \mathcal{H}_{\mathcal{S}_i}$ represent the quantum state of agent $i$ at time $t$, with each agent assigned its own Hilbert space $\mathcal{H}_{\mathcal{S}_i}$. The joint state for $N$ agents is represented as $\ket{\M{\Psi}_t} = \ket{\M{\psi}_{1, t}} \otimes \ket{\M{\psi}_{2, t}} \otimes \dots \otimes \ket{\M{\psi}_{N, t}} \in \mathcal{H}_{\mathcal{S}}$. 

Each agent $i$ selects actions $\ket{\M{a}_{i, t}} \in \mathcal{H}_{\mathcal{A}_i}$ based on its policy $\B{U}_{\Theta_i}$:
\begin{align}
    \ket{\M{a}_{i, t}} = \B{U}_{\Theta_i} \ket{\M{\psi}_{i, t}}
\end{align}

The joint transition operator $\hat{T}_{\text{joint}}$ governs the evolution of the joint state $\ket{\M{\Psi}_{t+1}}$ based on the combined actions:
\begin{align}
    \ket{\M{\Psi}_{t+1}} = \hat{T}_{\text{joint}} \left(\ket{\M{\Psi}_t}, \ket{\M{a}_{1, t}}, \dots, \ket{\M{a}_{N, t}}\right)
\end{align}

A joint reward operator $\hat{R}_{\text{joint}}$ evaluates the collective actions, facilitating both cooperative and competitive strategies:
\begin{align}
    r_{t} = \langle \M{\Psi}_t | \hat{R}_{\text{joint}} | \M{\Psi}_t \rangle
\end{align}

Each agent uses a replay memory to store past experiences, represented as tuples $(\ket{\M{\psi}_{i, t}}, \ket{\M{a}_{i, t}}, r_t, \ket{\M{\psi}_{i, t+1}})$. This replay memory enables the agent to sample experiences for training, which helps to break temporal correlations and improve learning stability. The policy $\Theta_i$ for each agent is optimized using a classical optimizer, where the loss functions are defined as follows:

\begin{itemize}
    \item \textbf{Actor Loss} ($L_a$) aims to maximize the expected value of the critic's Q value: \begin{align}
        L_{a, i} = -Q_{\Theta_{c, i}}(v_{\pi_i})
    \end{align}

    \item \textbf{Critic Loss} ($L_c$) minimizes the difference between the predicted Q-value and the target Q-value, defined in eq.~\ref{eq:loss}. 
\end{itemize}

\begin{figure*}[b] 
\vspace{-0.2cm}
\hrulefill 
\begin{equation}
    L_{c, i} = \left( r_t + \gamma Q_{\Theta_{c, i}}(\ket{\M{\psi}_{i, t+1}}, \ket{\M{a}_{i, t+1}}) 
    - Q_{\Theta_{c, i}}(\ket{\M{\psi}_{i, t}}, \ket{\M{a}_{i, t}}) \right)^2
    \label{eq:loss}
\end{equation}

\end{figure*}

Each agent optimizes its policy parameters $\Theta_i$ to maximize the expected cumulative reward, accounting for interdependencies with other agents through the shared quantum environment. The combination of replay memory and loss-based optimization helps stabilize and enhance the training process for each agent within this multi-agent quantum reinforcement learning framework.

QMARL is an emerging research area; for example, \cite{yun2022quantum} proposes a centralized-training, decentralized-execution framework using variational quantum circuits, which shows significant reward gains over classical MARL baselines under NISQ constraints. This work was later extended to a meta-learning setting through Quantum Multi-Agent Meta Reinforcement Learning \cite{yun2023quantum}. More recently, \cite{park2023quantum} introduced Entangled Quantum Multi-Agent Reinforcement Learning (eQMARL). The eQMARL is a distributed quantum actor–critic framework that facilitates agent cooperation through quantum entanglement. The proposed system uses a split quantum critic connected across agents via a quantum channel, eliminating the need for local observation sharing and reducing classical communication overhead.

\subsubsection{Free energy-based reinforcement learning}

The free energy-based reinforcement learning (FERL) draws on statistical physics, using free energy to guide learning. In a quantum context, \ac{FERL} models the state distribution of the environment with quantum Boltzmann machines, where the landscape of free energy informs policy optimization. The policy $\B{U}_\Theta$ is adjusted to minimize the free energy $F$, defined by:
\begin{align}
    F = -\frac{1}{\beta} \log \left( \sum_{\M{\psi}} e^{-\beta E(\M{\psi})} \right)
\end{align}
where $\beta$ is the inverse temperature and $E(\M{\psi})$ represents the energy of state $\M{\psi}$. By sampling states with low free energy, the agent explores favorable states, adjusting its policy to optimize cumulative rewards in complex environments.

Several papers have investigated this Boltzmann-machine approach to QRL, including demonstrations of FERL on quantum hardware \cite{levit2017free}, as well as hybrid actor–critic algorithms in which the actor is classical and the critic is implemented with a quantum Boltzmann machine \cite{schenk2024hybrid}.

\subsubsection{Quantum variational autoencoder for reinforcement learning}
The Quantum variational autoencoder for reinforcement learning (QVARL) compresses high-dimensional quantum states into lower-dimensional latent spaces, improving learning efficiency. The quantum autoencoder, parameterized by $\Theta$, encodes a state $\ket{\M{\psi}}$ into a latent state $\ket{\M{z}}$ to reduce complexity:
\begin{align}
    \ket{\M{z}} = \B{U}_{\Theta_{\text{enc}}} \ket{\M{\psi}}
\end{align}
The policy $\B{U}_{\Theta_{\text{policy}}}$ then operates in this reduced space, simplifying the agent’s learning process:
\begin{align}
    \ket{\M{a}} = \B{U}_{\Theta_{\text{policy}}} \ket{\M{z}}
\end{align}
This method enhances convergence and performance, especially in large or continuous state spaces. In some implementations, such as \cite{nagy2024hybrid}, the autoencoder that produces the latent representation is classical, while the policy network is operating on this latent space is quantum (quantum agent).

\subsubsection{Quantum Hierarchical Reinforcement Learning}

Quantum hierarchical reinforcement learning (QHRL) extends the concept of hierarchical policy learning to quantum environments, 
where complex tasks are decomposed into interdependent subtasks. 
A high-level quantum policy $\B{U}_{\Theta_{\text{high}}}$ defines abstract sub-goals or meta-actions for a given state $\ket{\M{\psi}}$:
\begin{align}
    \ket{\M{a}_{\text{high}}} = \B{U}_{\Theta_{\text{high}}} \ket{\M{\psi}},
\end{align}
while a low-level policy $\B{U}_{\Theta_{\text{low}}}$ executes these sub-goals through specific quantum actions:
\begin{align}
    \ket{\M{a}_{\text{low}}} = \B{U}_{\Theta_{\text{low}}} \ket{\M{a}_{\text{high}}}.
\end{align}
This hierarchical structure allows layered decision-making, 
where quantum policies at different abstraction levels cooperate to improve learning stability and task efficiency.
Recent work \cite{ZMZW:24:CIS} 
demonstrated a two-level QHRL framework for relation extraction tasks, 
highlighting that hierarchical quantum policies can effectively decompose complex objectives 
and enhance learning performance—an idea that generalizes naturally to quantum reinforcement learning settings.

\section{QRL Algorithms}\label{sec:Algorithms}

By leveraging quantum principles, QRL algorithms extend classical reinforcement learning, aiming for potential speedups or enhanced performance in complex environments. Broadly, these algorithms can be categorized into two main types:

\begin{itemize}
    \item \textit{Policy-Based Methods:} Aim to directly learn an optimal policy that maps states to actions, without necessarily relying on an intermediate value function, like policy gradient methods.
    \item \textit{Value-Based Methods:} Focus on learning an optimal value function that estimates the expected long-term return of taking specific actions in given states, like Q-learning methods.
\end{itemize}

In practice, it is also possible to combine the strengths of both approaches in so-called Actor-Critic methods. Here, the actor (policy-based component) learns the policy directly, while the critic (value-based component) estimates a value function to guide and stabilize the actor’s updates. 

In this section, we discuss the main QRL algorithms that have been explored in the literature. Specifically, we will discuss \textit{quantum policy gradient}, \textit{quantum Q-learning}, and \textit{quantum actor-critic}. For each algorithm, we will provide a short tutorial. Table~\ref{tab:QRLalgorithms} provides a comparison between these algorithms.

\begin{table*}[ht]
    \centering
    \caption{Comparison of Main QRL Algorithms}
    \label{tab:QRLalgorithms}
    \begin{tabular}{ccccc}
    \toprule
        \textbf{Algorithm} & \textbf{learning method }& \textbf{optimize} & \textbf{On-/Off-policy} &\\
        \midrule
        Quantum policy gradient & Monte Carlo (for REINFORCE)  & policy & on-policy &\\
        Quantum Q-learning & Temporal Difference (TD) & value function & off-policy &\\
        Quantum Actor-critic & Temporal Difference (TD) & policy$+$value function & usually on-policy &\\
        \bottomrule
    \end{tabular}
    \\[1ex]

\end{table*}

\subsection{Quantum policy gradient}
Quantum policy gradient methods optimize the policy parameters $ \Theta $ by directly computing the gradient of the expected cumulative reward $ \mathbb{E}[R_t] $ with respect to $ \Theta $, using the policy $ \B{U}_\Theta $ for action selection. The update rule is the same in eq.~\ref{eq:update}. Several implementations of quantum policy gradient have been explored in the literature \cite{meyer2023quantum}, \cite{meyer2023quantum2} and \cite{sequeira2023policy}.

One possible implementation of quantum policy gradient methods is proposed by \cite{jerbi2021parametrized}. Below, we include a short tutorial on their approach, which uses parameterized quantum circuits as the policy model and applies the REINFORCE algorithm to optimize its parameters. Their method introduces two policy variants: RAW-PQC and SOFTMAX-PQC.

In classical reinforcement learning, the policy is typically modeled as a neural network. In contrast, \cite{jerbi2021parametrized} uses a parameterized quantum circuit that takes the state $s$ as input and prepares the quantum state $|\psi_{s,\Theta}\rangle $, where $\Theta$ denotes trainable parameters. From this quantum state, the agent either measures the state and directly maps the outcome to an action (RAW-PQC), or computes observables for each action and applies a softmax to obtain action probabilities (SOFTMAX-PQC).

The parameterized quantum circuit used in this method follows a hardware-efficient architecture consisting of alternating layers, where the encoding layers and the variational layers are applied in an alternating fashion. The encoding layers consist of single-qubit rotations $R_z$ and $R_y$ to embed the input state into the circuit. The variational layers also include single-qubit rotations  $R_z$ and $R_y$, along with entangling gates such as controlled-Z (CZ) gates. This PQC architecture is used for both the RAW-PQC and SOFTMAX-PQC policy variants.


\begin{enumerate}
\item \textit{RAW-PQC:}
In RAW-PQC, the Hilbert space is partitioned into regions corresponding to each possible action available to the agent. In other words, each action $a \in \mathcal{A}$ is associated with a projector $P_a$; together, these projectors partition the Hilbert space. The probability of selecting action $a$ is: 
\begin{align}
\pi_\Theta(a \mid s) = \langle \psi_{s,\Theta} | P_a | \psi_{s,\Theta} \rangle
\end{align}
The gradient for updating the parameters $\Theta$ was derived as: 
 \begin{align}
\nabla_\theta \log \pi_\theta(a \mid s) = \frac{\nabla_\theta \langle P_a \rangle_{s, \theta}}{\langle P_a \rangle_{s, \theta}}
\end{align}
While RAW-PQC is simple and leverages the inherent probabilistic nature of quantum measurement to select actions, it lacks a mechanism to directly control the degree of exploration versus exploitation. In other words, there is no tunable parameter that allows the agent to adjust how greedy or exploratory its behavior should be. As training progresses, the action probabilities often increase sharply around a single outcome, which can reduce variability in action selection and limit exploration during evaluation.

\item \textit{SOFTMAX-PQC:}
To address this limitation, a non-linear activation function (i.e softmax) is applied to the expectation values $\langle \psi_{s,\Theta} | P_a | \psi_{s,\Theta} \rangle$. This variant, known as SOFTMAX-PQC, introduces a temperature parameter $\beta$ that is adjustable, which allows the agent to control the greediness of the policy. Here, the projections $P_a$ are generalized to arbitrary trainable Hermitian operators $O_a$, which are associated with each action, and are given by:
\begin{align}
O_a = \sum_i w_{a,i} H_{a,i}
\end{align}
where $w_{a,i}$ are trainable weights. The policy of SOFTMAX-PQC can be defined as: 
\begin{align}
\pi_\theta(a \mid s) = \frac{e^{\beta \langle O_a \rangle_{s, \theta}}}{\sum_{a'} e^{\beta \langle O_{a'} \rangle_{s, \theta}}}
\end{align}
where the expectation value is given by $\langle O_a \rangle_{s, \theta} = \langle \psi_{s, \boldsymbol{\phi}, \boldsymbol{\lambda}} | \sum_i w_{a,i} H_{a,i} | \psi_{s, \boldsymbol{\phi}, \boldsymbol{\lambda}} \rangle$, where $\Theta$ includes all trainable parameters. The gradient of this policy is given by: 
\begin{align}
\nabla_\theta \log \pi_\theta(a \mid s) = \beta ( 
\nabla_\theta \langle O_a \rangle_{s, \theta}   \nonumber \\
- \sum_{a'} \pi_\theta(a' \mid s) \nabla_\theta \langle O_{a'} \rangle_{s, \theta} 
)
\end{align}
\end{enumerate}

To train the circuits for both policy variants, the Monte Carlo policy gradient algorithm REINFORCE is used. The agent maximizes the expected return by updating the circuit parameters $\Theta$ via gradient ascent.

\subsection{Q-learning using variational quantum algorithms}

Unlike the policy-gradient approach, which directly optimizes the policy, deep Q-learning uses a parameterized quantum circuit to estimate the agent's Q-function. Similarly to how a neural network approximates Q-values in classical deep Q-learning, a parameterized quantum circuit takes on this role in its quantum counterpart, enabling the agent to infer its policy by selecting actions that maximize the estimated Q-values. This builds on the classical Q-learning algorithm introduced in Section~\ref{sec:Pre}, where the agent updates a Q-table based on the Bellman optimality equation.

Several papers have explored the use of parameterized quantum circuits as a value function approximator \cite{chen2023quantum,chen2020variational,chen2024deep}. In \cite{chen2023quantum3}, Quantum Deep Recurrent Q-Learning (QDRQN) was proposed, where a quantum long short-term memory (QLSTM) network was integrated into the deep Q-learning framework to serve as a Q-value estimator. In the following, we provide a tutorial on a specific implementation of quantum Q-learning from \cite{skolik2022quantum}.

In \cite{skolik2022quantum}, parameterized quantum circuits are used to approximate the Q-function. The classical neural network used in deep Q-learning is replaced by a variational quantum circuit that maps input states to Q-values corresponding to each possible action.

In this approach, the classical neural network is replaced by a parameterized quantum circuit. The ansatz used is hardware-efficient, making it highly expressive. Here, ``expressive'' denotes the ability of the parameterized quantum circuit to span a high-dimensional subspace of the Hilbert space, allowing it to approximate complex quantum transformations required for policy learning.
 Each layer of the PQC consists of single-qubit rotations $R_y$ and $R_z$ followed by a series of controlled-Z (CZ) entangling gates arranged in a daisy chain pattern. The circuit takes an environment state as input and outputs Q-values corresponding to each available action. To encode classical states into the quantum circuit, $R_x$ gates are applied. Depending on whether the environment has discrete or continuous state spaces, different preprocessing strategies are used. For discrete states, basis encoding is used. For continuous states, input components $x_i$ are first scaled using an $\arctan$ function to map them into the range $[-\frac{\pi}{2}, \frac{\pi}{2}]$.

To enhance the expressivity of the circuit, two techniques are introduced. First, \textit{data re-uploading} can be used, in which layers of data encoding and variational gates are repeated in an alternating fashion. Second, trainable weights $w_d$ can be applied to the input data, allowing the model to learn the appropriate input scaling. In this case, the scaled input becomes:
\begin{align}
x_i' = \arctan(x_i \cdot w_{d,i}).
\end{align}

Q-values for each action are computed as expectation values of an observable on the quantum state prepared by the PQC:
\begin{align}
Q(s, a) = \langle 0^{\otimes n} | U_\theta^\dagger(s) \, O_a \, U_\theta(s) | 0^{\otimes n} \rangle
\end{align}
where $U_\theta(s)$ is the Q-network PQC with and encoded state $s$ and parameter $\theta$, while $n$ is the number of qubits.
A problem that arises is that Quantum observables have fixed ranges, and Q-values can be arbitrarily large. Therefore, the output was made scalable with trainable weights
\begin{align}
Q(s, a) = \langle 0^{\otimes n} | U_\theta(s)^\dagger O_a U_\theta(s) | 0^{\otimes n} \rangle \cdot w_{o_a}
\end{align}

During training, the agent interacts with the environment to generate experience tuples $(s_t, a_t, r_{t+1}, s_{t+1})$, where $s_t$ is the current state, $a_t$ is the action taken, $r_{t+1}$ is the immediate reward in the next state, and $s_{t+1}$ is the next state. These transitions are stored in an experience replay buffer, from which minibatches $\mathcal{B}$ are drawn uniformly at random to remove temporal correlations, as in classical deep Q-learning.  

A target network with parameters $\theta'$ periodically updated from the main network $\theta$ is used to compute the bootstrap target. The loss function is:  
\begin{align}
\mathcal{L}(\theta) 
&= \frac{1}{|\mathcal{B}|} 
   \sum_{(s,a,r,s') \in \mathcal{B}} 
   \Big( Q_{\theta}(s,a)  \nonumber \\[2pt]
&\quad - \big[\, r + \max_{a'} Q_{\theta'}(s',a') \,\big] \Big)^{2}.
\label{eq:dqn-loss}
\end{align}

\subsection{Quantum actor-critic}
The quantum actor-critic method uses two components: the actor, which updates the policy parameters $ \Theta $ in the policy $ \B{U}_\Theta $, and the critic, which estimates the quantum value function $ V(\ket{\M{\psi}}) $. The actor updates $ \Theta $ based on feedback from the critic’s value function estimate:
\begin{align}
    \Theta \leftarrow \Theta + \eta \nabla_\Theta \mathbb{E}\left[ R_t \mid V(\ket{\M{\psi}}) \right]
\end{align}

Several quantum actor-critic implementations have been studied in the literature, such as in \cite{chen2023asynchronous} and \cite{lan2021variational}. Recent work extended the quantum actor-critic framework by integrating Quantum Long Short-Term Memory (QLSTM), e.g., in \cite{chen2024efficient}, and by combining quantum actor-critic with fast weights, as demonstrated in \cite{chen2024learning}. In some actor-critic implementations, the critic itself does not have to be quantum; instead, a classical neural network is often used to approximate the value function \cite{park2023quantum}. This hybrid setup allows the actor to leverage quantum expressivity, while the critic benefits from the stability and efficiency of classical function approximation. In what follows, we present a short tutorial on the quantum actor-critic method, based on \cite{kwak2021introduction}, to illustrate how these components work together in practice.

The quantum actor in this method is implemented using a VQC. Each component of the environment state is encoded using single-qubit rotations $R_y$. The circuit then applies $R_x$, $R_y$ and $R_z$ on all qubits in the circuit, followed by controlled-Z gates. All qubits are then measured and a softmax function is applied to map the expectation values to actions. The circuit outputs action-values. 

While the actor is quantum, the critic is kept classical and is responsible for evaluating the decisions taken by the policy using the value functions. The critic is implemented as a feedforward neural network that estimates the state-value function.

In order to train the critic to evaluate the actor, a reply buffer needs to be used, where it stores experience as tuples $(s_t, a_t, R_{t+1}, s_{t+1})$. After collecting experience, mini batch are sampled from the buffer and are used to calculate the temporal difference, which is the critic target:
\begin{align}
y_j &=
\begin{cases}
R_j, 
& \text{if } s_{j+1} \text{ is terminal}, \\[6pt]
R_j + \gamma \max_{a'} Q(s_{j+1}, a'; \theta), 
& \text{otherwise.}
\end{cases}
\label{eq:target}
\end{align}

The critic is then trained to make $V(s)$ match $y_j$, $\delta_j = y_j - V(s_j)$. While critic is training, in parallel, the actor is also being evaluated by the critic. This is done using the advantage function $A_t$. The actor is trained using Proximal Policy Optimization (PPO), and the loss function becomes: 

\begin{align}   
\mathcal{L}_{t}(\theta) = \min \Big( r_t(\theta) A_t,\; 
\text{clip}\big(r_t(\theta),\, 1 - \epsilon,\, 1 + \epsilon \big) A_t \Big).
\end{align}

\begin{figure*}[b] 

\hrulefill

\begin{equation}
    V(\ket{\M{\psi}_t}) \leftarrow V(\ket{\M{\psi}_t}) + \\ \eta \left[ r_t + \gamma V(\ket{\M{\psi}_{t+1}}) - V(\ket{\M{\psi}_t}) \right]
    \label{eq:TD}
\end{equation}

\end{figure*}

\begin{table*}[ht]
\centering
\caption{Quantum Reinforcement Learning Frameworks \acp{SDK}}
\begin{tabularx}{\textwidth}{l l c c c l}
\toprule
\textbf{Framework} & \textbf{ML Integration} & \textbf{GPU acceleration} & \textbf{QRL Tools} & \textbf{Release / First Commit} & \textbf{Hardware Backend} \\
\midrule
\href{https://github.com/Qiskit/qiskit}{Qiskit} & PyTorch, Singularity & $\checkmark$ & $\times$ & Mar 2017 & IBM Devices, Simulators \\
\href{https://github.com/quantumlib/Cirq}{Cirq} & TensorFlow & $\times$ & $\times$ & Jul 2018 & Google Devices, Simulators \\
\href{https://github.com/PennyLaneAI/pennylane}{PennyLane} & PyTorch, TensorFlow, JAX & $\checkmark$ & $\times$ & Oct 2018 & Multiple Quantum Devices \\
\href{https://github.com/tensorflow/quantum}{TensorFlow Quantum} & TensorFlow & $\times$ & $\times$ & Mar 2020 & Simulated Quantum Circuits \\
\href{https://github.com/mit-han-lab/torchquantum}{TorchQuantum} & PyTorch & $\checkmark$ & $\times$ & Apr 2022 & Simulated Quantum Circuits \\
\href{https://github.com/NVIDIA/cuda-quantum}{CUDA Quantum} & C++, Python & $\checkmark$ & $\times$ & Aug 2023 & NVIDIA GPUs, Simulated QPUs \\
\href{https://github.com/sQUlearn/squlearn}{sQUlearn} & scikit-learn & $\checkmark$ & $\checkmark$ & May 2023 & Multiple Quantum Devices \\
\href{https://github.com/qlan3/QuantumExplorer}{QuantumExplorer} & PyTorch, PennyLane & $\checkmark$ & $\checkmark$ & Jun 2023 & Simulated Quantum Circuits \\
\href{https://github.com/QuTech-Delft/qgym}{qgym} & OpenAI Gym & $\times$ & $\checkmark$ & Sep 2023 & Simulated Quantum Compilers \\
\href{https://github.com/Sampreet/quantrl}{Quantrl} & Custom & $\checkmark$ & $\checkmark$ & Oct 2023 & Simulated Quantum Systems \\

\bottomrule
\end{tabularx}
\label{tab:SDK}
\end{table*}

\section{QRL Benchmarking} \label{sec:benchmark}

While QRL is a rapidly growing field, it is currently experiencing significant problems with benchmarking. The field lacks a unified benchmark and standardized metrics, making it difficult to properly evaluate and compare different algorithms \cite{kruse2025benchmarking}.
Claiming that \textit{“algorithm A outperforms algorithm B”} in QRL is challenging due to its high sensitivity to hyperparameters and multiple sources of randomness. Small changes in learning rate, circuit depth, or number of qubits can drastically alter results. Furthermore, QRL faces multiple additional sources of randomness, such as hardware noise, which can vary from one device to another, and randomness in shots from measurement, all of which hinder consistent evaluation. Additional factors such as weight initialization, action sampling, and environment stochasticity further reduce reproducibility, making fair comparisons between algorithms difficult.
Besides noise, the environment plays a crucial role in establishing whether a quantum algorithm truly outperforms its classical counterpart. The environment must be sufficiently complex to challenge classical algorithms, yet structured in a way that leverages the unique strengths of quantum computation. Striking this balance is difficult, making environment design choices a significant challenge. 

In response to these challenges, recent efforts have emerged to establish more rigorous and standardized benchmarking practices in QRL, marking the first steps toward a more reliable and comparable evaluation landscape. The authors in \cite{meyer2025benchmarking} proposed a new benchmarking method, which evaluates the sample complexity (i.e., the amount of interactions between the agent and the environment to achieve a certain performance) of heuristic algorithms using a statistical estimator. They also introduced a new benchmarking environment that has adjustable levels of complexity. 
Similarly, the authors in \cite{kruse2025benchmarking} introduced a series of metrics used to evaluate QRL algorithms: Performance, sample efficiency, number of circuit executions, quantum clock time and qubit scaling. These metrics go beyond traditional RL evaluation (performance and sample efficiency) by incorporating quantum-specific considerations.   
Finally, in \cite{ikhtiarudin2025benchrl}, the authors propose a weighted ranking metric that incorporates accuracy, circuit depth, gate count, and computational efficiency, enabling fair comparisons in quantum architecture search tasks.

\begin{figure}[htb]
    \centering
    \includegraphics[width=\linewidth]{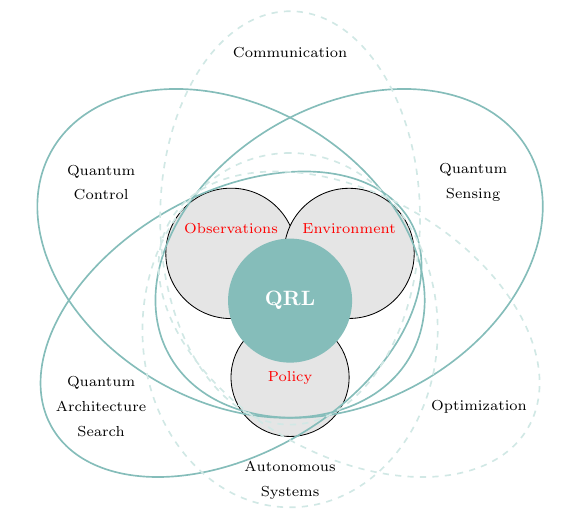}
    \caption{Illustration of \ac{QRL}'s transversal applicability, showcasing its potential to enhance learning and decision-making in both quantum-specific and classical application domains.}
    \label{fig:app}
\end{figure}


\section{RL Applications}\label{sec:classicalRLappl}
Classical reinforcement learning has also been employed to address tasks where the application itself is quantum, such as quantum control, quantum error correction, quantum architecture search, quantum sensing and quantum key distribution. In this section, we briefly survey recent advances in these areas, highlighting how purely classical agents and algorithms can optimize the behavior of quantum systems despite operating on classical hardware.

\subsection{Quantum Control}
Recent advancements in quantum computing have shifted the focus from merely increasing qubit counts to enhancing qubit quality through error correction. Concurrently, the transition from pulse-level control to fractional gates is streamlining quantum operations, reducing circuit depth, and improving efficiency \cite{ibm_fractional_gates}.  These developments underscore the critical role of sophisticated quantum control techniques in achieving reliable and scalable quantum computation. Quantum control involves manipulating quantum systems to achieve specific objectives, such as state transitions or implementing quantum operations \cite{PEHM:22:Quantum, ZXZCW:24:arXiv, JPWGD:22:PRA,GST:24:PRA, AZ:19:EPL}. This is accomplished by applying external fields, such as lasers or magnetic fields, to influence the system's Hamiltonian, which governs its evolution \cite{DP:2010:IET_CTA}. Mathematically, for a state $\ket{\psi(t)}$, the evolution is governed by the time-dependent Hamiltonian:
\begin{align}
H(t) = H_0 + \sum_i u_i(t) H_i,
\end{align}
where $u_i(t)$ are control parameters. The system evolves as:
\begin{align}
\ket{\psi(t)} = U(t, 0) \ket{\psi(0)},
\end{align}
with:
\begin{align}
U(t, 0) = \mathcal{T} \exp\left(-\frac{i}{\hbar} \int_0^t H(t') dt'\right).
\end{align}
The objective is to optimize $u_i(t)$ to maximize a performance metric such as fidelity:
\begin{align}
F = |\braket{\psi_\text{target}|\psi(t)}|^2.
\end{align}
\ac{RL} can automate the optimization of control parameters by treating the system's state as the environment, control actions as the RL agent's actions, and fidelity as the reward\cite{LWFGZ:22:SoftComp}. The reward function, which reflects the optimization objective, can be expressed as:
\begin{align}
C = \chi (1 - F[\mathbf{U}(T)]) + \beta L_\text{tot} 
\\ + \mu \int_{0}^{T} \left[ g^2(t) + f^2(t) \right] dt + \kappa T,
\end{align}
where $C$ represents the total cost function to be minimized. $F[\B{U}(T)]$ is the fidelity, which quantifies the overlap between the evolved state and the target state at the final time $T$. $L_\text{tot}$ denotes the total leakage error, which captures the probability of the system deviating from its intended Hilbert space. $g(t)$ and $f(t)$ are control parameters representing system constraints, such as amplitude and frequency limits of the applied controls, and $\mu$ is a penalty coefficient associated with these constraints. $\kappa$ is the penalty weight for the total runtime $T$, and $\chi, \beta, \mu, \kappa$ are hyperparameters balancing the contributions of fidelity, leakage, control effort, and runtime in the cost function.

An RL agent can learn a policy $\pi(\M{s}_t)$, which maps states $\M{s}_t$ of the quantum system to control actions $\M{a}_t$ to maximize the cumulative discounted reward. The reward is defined as:
\begin{align}
J(\pi) = \mathbb{E} \left[ \sum_{t=0}^T \gamma^t R_t \right],
\end{align}
where $R_t$ is the reward at time $t$, and $\gamma$ (with $0 < \gamma \leq 1$) is the discount factor that determines the importance of future rewards. The policy is iteratively optimized using techniques like policy gradient, which updates the policy parameters $\theta$ as:
\begin{align}
\nabla_\theta J(\pi_\theta) = \mathbb{E}_\pi \left[ \nabla_\theta \log \pi_\theta(\M{a}_t | \M{s}_t) R_t \right],
\end{align}
where $\pi_\theta(\M{a}_t | \M{s}_t)$ is the probability of taking action $\M{a}_t$ given the state $\M{s}_t$ under the current policy $\pi_\theta$. This approach allows the agent to identify control trajectories $\{u_i(t)\}$ that minimize the total cost function $C$ while considering system constraints and noise, achieving efficient and high-fidelity quantum control \cite{NBSN:19:npj_QI,MDDC:22:IEEE_J_NNLS}. 

\ac{RL} has been widely applied in many use cases for quantum control. For instance, the authors in \cite{BDSWPM:18:PRX} demonstrate how RL can optimize quantum control protocols across different system phases, revealing phase transitions in the control landscape and offering a model-free, scalable approach for high-fidelity state transitions in complex quantum systems. Similarly, \cite{KVPDFC:24:arXiv, WCDYCHZXDL:24:arXiv} introduces an RL-based approach to optimize quantum circuit transpilation, achieving near-optimal synthesis for various circuit types and significant reductions in gate depth and count, outperforming traditional heuristic and optimization methods in efficiency and scalability. Additionally, Qubit routing, formulated as an \ac{RL} problem, uses deep Q-learning to minimize SWAP gate overhead by optimizing dynamic qubit permutations, significantly improving circuit depth and hardware efficiency \cite{PHSJM:22:ACM_TQC, HS:18:arXiv}. In \cite{RLFOBANRKBB:23:NC}, a real-time reinforcement learning agent is implemented on an FPGA for low-latency quantum feedback, achieving high-fidelity control and initialization of superconducting qubits without requiring explicit system models. Although the quantum control problem has been effectively mapped to an \ac{RL} problem, quantum-inspired exploration strategies and reward schemes have shown superior performance compared to traditional \ac{RL}  methods in scenarios such as one-qubit, two-level open systems, and many-qubit systems, showcasing enhanced stability, efficiency, and learning capabilities under experimental constraints \cite{YZC:24:IEEE_J_CST}.

\subsection{Quantum Error Correction}
Quantum computers are inherently susceptible to noise and decoherence, making errors during computation unavoidable. To achieve reliable large-scale quantum computers, Quantum Error Correction (QEC) is therefore required. However, implementing QEC is far more complicated than classical error correction, where redundancy can be easily used by simply copying bits. In the quantum domain, there are three challenges \cite{roffe2019quantum}:

\begin{itemize}
    \item \textit{No-cloning of quantum states:} In classical codes, data can be duplicated to achieve redundancy, but the quantum no-cloning theorem forbids making identical copies of an unknown state.
    \item \textit{Multiple error types:} Classical bits suffer only bit-flip errors, while qubits are vulnerable to both bit-flips (X errors) and phase-flips (Z errors), requiring codes that correct both simultaneously.
    \item \textit{Measurement induced collapse:} Classical bits can be read without disturbance, but measuring a qubit can destroy the encoded information. 
\end{itemize}

Therefore, QEC relies on carefully engineered encodings and control strategies to detect and correct errors without disturbing the stored information. While traditional QES schemes such as surface codes, stabilizer codes, and other established quantum codes are powerful, they usually use a large number of qubits and more complex optimization \cite{veeresh2024rl}. Building on these principles, researchers have explored RL methods to automate and optimize QEC.

For example, Deep RL is used for quantum error correction on the toric code under uncorrelated bit-flip or phase-flip noise \cite{andreasson2019quantum}. This was done by training an agent to find near-optimal correction paths and achieving accuracy comparable to a Minimum-Weight Perfect Matching decoder. 

The task of decoding fault-tolerant surface codes can likewise be reformulated as a sequential decision-making problem, where a learning agent interacts with the code’s syndrome data \cite{sweke2020reinforcement}.
In this framework, the decoder behaves as an RL agent, receiving observations from the quantum code environment and selecting corrective actions to reduce logical errors. Using a deep-Q learning approach, they train classical networks that successfully learn high-performance decoding strategies under realistic noise conditions.

RL has also been used to design autonomous quantum error correction (AQEC) protocols. An RL agent identified optimal bosonic codewords for AQEC in superconducting systems, achieving high-fidelity protection of logical qubits \cite{zeng2023approximate}. 

RL has also been used to directly target bit-flip and depolarizing noise in surface-code architectures \cite{veeresh2024rl}, where agents are trained to lower bit-flip errors by analyzing error rates and monitoring qubit lifetimes.

low-weight quantum error correcting codes with dramatically reduced physical qubit overhead have also been discovered, by applying a Proximal Policy Optimization agent to stabilizer codes \cite{he2025discovering}.

\subsection{Quantum Architecture Search}\label{QAS}
\Ac{QAS} automates the design of quantum circuit architectures tailored for specific applications and hardware constraints. It searches the space of possible configurations to identify efficient architectures in terms of depth, gate fidelity, and overall performance \cite{MJP:24:arXiv, ZPP:22:ACSYS}. Inspired by classical neural architecture search, \ac{QAS} addresses unique quantum challenges, including unitary constraints, noise sensitivity, and hardware-specific limitations \cite{DHYHT:22:npj_QI}. The efficiency of \acp{VQC} depends heavily on the architectures used due to:
\begin{itemize} 
\item \textit{Expressivity:} The architecture determines the \ac{VQC}'s ability to represent the target solution space. 
\item \textit{Trainability:} Poorly designed circuits can lead to barren plateaus, where gradients vanish and training becomes infeasible. 
\item \textit{Hardware Compatibility:} Constraints like limited qubit connectivity and gate fidelities require customized architectures for efficient execution. \end{itemize}

\Ac{QAS} optimizes expressivity, trainability, and hardware compatibility, addressing the exponential growth of the search space with circuit size. Mathematically,\ac{QAS} seeks to find the optimal quantum circuit architecture $\mathcal{A}^*$ and parameters $\vec{\M{\theta}}^*$ to minimize a cost function $\mathcal{L}$:
\begin{align}
(\mathcal{A}^*, \vec{\M{\theta}}^*_{\mathcal{A}^*}) = \arg\min_{\mathcal{A}, \vec{\M{\theta}}_{\mathcal{A}}} \mathcal{L}(\mathcal{A}, \vec{\M{\theta}}_{\mathcal{A}})
\end{align}

To efficiently explore the large architecture space, QAS can be cast as a reinforcement learning problem, where an RL agent builds and evaluates circuits to discover high-performing designs \cite{kuo2021quantum}. In this RL framework:
\begin{itemize}
    \item \textbf{State Space:} $\M{s}_t$ represents the current circuit configuration.
    \item \textbf{Action Space:} $\M{a}_t$ modifies the circuit (e.g., adding gates or parameters).
    \item \textbf{Policy:} $\pi_\B{\Theta}(\M{a}_t \mid \M{s}_t)$ maps states to actions, parameterized by $\B{\Theta}$.
    \item \textbf{Reward:} $R(\M{s}_T)$ evaluates the circuit's performance.
\end{itemize}

The \ac{RL} agent optimizes $\pi_\B{\Theta}$ to maximize expected cumulative rewards:
\begin{align}
\B{\Theta}^* = \arg\max_\B{\Theta} \mathbb{E}\left[\sum_{t=0}^{T} \gamma^t R(\M{s}_T) \mid \pi_\B{\Theta}\right]
\end{align}

Building on the foundations of \ac{QAS}, recent advancements underscore the critical role of \ac{RL} in optimizing quantum circuit architectures. For example, \cite{dai2024quantum} employs RL to automatically design and refine \ac{QML} models. Other frameworks, such as \textit{QAS-Bench} \cite{LPYSWY:23:ICML} and differentiable approaches  \textit{QuantumDARTS} \cite{WYL:23:ICML} illustrate how systematic evaluation and gradient-based optimization techniques enhance circuit exploration and performance. \ac{RL}, in particular, has demonstrated exceptional efficacy in hardware-constrained environments. A notable example is the \textit{Nearest-Neighbor Compilation} framework\cite{LLLL:23:QIP}, where \ac{RL} methods are employed to minimize the number of SWAP gates and reduce circuit depth, addressing key practical limitations.

Advanced \ac{RL} techniques, including \textit{RNN-policy gradient methods}\cite{WWF:24:QIP} and \textit{recursive RL for \ac{QAOA}}\cite{PJBD:24:EPJ_QT}, further showcase \ac{RL}’s adaptability in sequential gate design and parameter optimization. These approaches achieve superior efficiency and faster convergence by dynamically navigating the complex design space of quantum circuits. Furthermore, \ac{RL}-driven frameworks such as \textit{KANQAS}\cite{KSS:24:arXiv} exemplify the power of hierarchical modeling to explore architecture spaces while addressing task-specific constraints efficiently.

By aligning circuit expressivity, trainability, and hardware compatibility, \ac{RL} not only automates and enhances the \ac{QAS} process but also fosters significant innovations in tailoring architectures for complex quantum tasks. As such, \ac{RL} has emerged as a pivotal tool for advancing the capabilities of \acp{VQC}, driving progress in algorithmic design and practical implementation.

\subsection{Quantum Sensing} 
Quantum sensing exploits quantum mechanical principles, such as superposition and entanglement, to achieve high-precision measurements of physical parameters like magnetic fields, time, and gravity. By leveraging the sensitivity of quantum states to external perturbations, quantum sensors surpass classical sensors in accuracy and efficiency \cite{SFB:20:NJP}.
The operation of a quantum sensor is governed by the evolution of its quantum state under a parameter-dependent Hamiltonian $H(\theta)$, where $\theta$ is the parameter to be estimated. The quantum state evolves as:
\begin{align}
    \rho(\theta) = \B{U}(\theta) \rho_0 \B{U}^\dagger(\theta)
\end{align}
where $\rho_0$ is the initial quantum state, and $\B{U}(\theta) = \exp(-i H(\theta) t)$ is the unitary evolution operator, with $t$ being the evolution time.
The precision in estimating $\theta$ is bounded by the quantum Cramér-Rao bound:
\begin{align}
    (\Delta \theta)^2 \ge \frac{1}{m F_Q[\rho(\theta), H]}
\end{align}
where $m$ represents the number of independent measurements, and $F_Q[\rho(\theta), H]$ is the quantum Fisher information, which quantifies the sensitivity of the state $\rho(\theta)$ to variations in $\theta$ \cite{XFZ:22:npj_QI}. Here, the \ac{RL} agent can learn a policy $\pi_\B{\Theta}(\M{a}_t \mid \M{s}_t)$ that maps the quantum system's state $\M{s}_t$ to an action $\M{a}_t$, such as applying control pulses or adjusting measurement settings. The agent aims to maximize the expected cumulative reward, which is defined in terms of the estimation precision or sensitivity of the quantum sensor. This reward, inversely proportional to the parameter estimation variance $(\Delta \theta)^2$, allows the agent to iteratively improve its strategy by updating the policy parameters $\B{\Theta}$ using methods such as policy gradients. This process incorporates feedback from quantum dynamics, enabling the discovery of optimal control strategies that enhance precision and robustness under noise and hardware constraints.

Recent advances in \ac{RL} for quantum sensing have highlighted its versatility and effectiveness. \ac{RL} has been shown to optimize quantum sensor dynamics, achieving more than an order-of-magnitude improvement in sensitivity by designing nonlinear control pulses that counteract decoherence \cite{SFB:20:NJP}. Similarly, a deep \ac{RL} framework for time-dependent parameter estimation was proposed, employing a geometrically inspired reward function and a time-correlated control ansatz to achieve robust, sample-efficient estimation under noisy and noise-free conditions \cite{XFZ:22:npj_QI}. In the context of Bayesian quantum sensing, an \ac{RL}-based experimental design framework outperformed traditional methods by using particle filtering to optimize adaptive sensing strategies \cite{BZM:23:arXiv}. 
Further advancements include the application of deep \ac{RL} to quantum multiparameter estimation, effectively addressing resource limitations and eliminating reliance on precise system models \cite{CVPPCCSOS:23:Adv_Phot, XLLWYW:19:npjQI}. Additionally, \ac{RL} has been used to design robust entanglement generation protocols tailored to various noise levels and system parameters, while \ac{RL}-based feedback control strategies have demonstrated superior performance in improving the precision of quantum metrology, outperforming conventional methods in dynamic quantum systems \cite{QZHL:22:NJP, FRTG:22:PRX}.

\subsection{Quantum Key Distribution}

Quantum Key Distribution (QKD) allows two parties to share a classical secret key securely by exploiting quantum‐mechanical principles such as the no-cloning theorem. Any eavesdropping attempt destroys the quantum states, allowing the two parties to detect and discard compromised keys. Nevertheless, QKD still faces significant challenges in resource allocation: the key generation rate decreases exponentially with distance, making it difficult to meet the high-traffic demands of modern applications \cite{takeoka2014fundamental}. Traditional allocation methods, such as shortest-path routing, concentrate requests on a few links, causing even a bigger congestion problem \cite{zheng2025efficient}. To overcome these limitations, recent work explores classical deep RL agents that dynamically allocate wavelengths, time slots, or key resources. 

The authors in \cite{seok2025deep} proposed a method that uses deep RL to tackle the resource provisioning problem in QKD networks. In their method, a classical RL agent was trained to dynamically allocate key resources and network pathways in response to varying demands and network conditions.

QKD lightpath requests need to be updated frequently, making routing and resource assignments (RRA) challenging. Therefore, a deep RL scheme has been proposed to tackle the RRA problem in QKD-secured optical networks \cite{sharma2023deep}.

Quantum key pools (QKPs) sit between adjacent QKD nodes to manage key resources, but dynamic traffic makes key generation and consumption unbalanced, causing service blocking, key overflow, and degraded security when keys remain too long in QKP. To address these challenges, an RL-based routing and key-resource assignment algorithm has been proposed in \cite{zuo2020reinforcement}, in which a deep Q-learning agent is trained to choose routing actions that keeps QKP's key level within that safe range.

\section{QRL Applications} \label{sec:QRLApp}
The growing body of QRL research demonstrates its versatility, with applications spanning autonomous systems, optimization, and communication. In this section, we review the current literature showcasing how QRL has been applied across these fields.
\paragraph{Autonomous Systems}
\ac{QRL} plays a pivotal role in advancing autonomous systems by enabling precise decision-making, efficient resource utilization, and robust control under dynamic and uncertain conditions. Through its integration of quantum computing with reinforcement learning, \ac{QRL} addresses computational and operational challenges that traditional methods struggle to overcome in real-time autonomous applications. The versatility of \ac{QRL} is demonstrated across a range of tasks, including:
\begin{itemize}
    \item \textit{Reusable Rocket Landing}: \ac{QRL}-based controllers significantly improve stability and adaptability during reusable rocket landings under turbulent conditions, such as wind disturbances. They achieve faster convergence and higher cumulative rewards, as demonstrated in \cite{KCP:24:IEEE_J_VT}, while meeting the computational constraints of onboard systems and outperforming classical methods such as Deep Q-Networks.

    \item \textit{Robot Navigation}: By utilizing \acp{VQC}, \ac{QRL} frameworks efficiently encode high-dimensional state representations, enabling autonomous robots to navigate complex environments with fewer computational resources. This approach has proven particularly effective in static navigation tasks where classical deep reinforcement learning methods fall short \cite{HHWK:24:IEEE_O_ACC}.

    \item \textit{Self-Driving Cars}: In collision-free navigation tasks, \ac{QRL} models such as Nav-Q combine quantum critics with classical dimensionality reduction techniques to enhance decision-making efficiency \cite{SMK:2023:arXiv}. These hybrid systems accelerate convergence and improve safety indices, making them highly suitable for real-world autonomous driving scenarios.

    \item \textit{Multi-Drone Mobility Control}: \ac{QRL}-based \ac{QMARL} frameworks optimize multi-drone coordination and task allocation in dynamic environments \cite{PKJK:24:APWCS}. These systems enable efficient policy learning, robust action planning, and stable performance, critical for applications such as surveillance and resource delivery.

    \item \textit{Pedestrian Interaction Modeling}: \ac{QRL}’s integration into Quantum-like Bayesian models enhances the prediction of pedestrian behaviors, addressing irrational and unpredictable actions in traffic scenarios  \cite{SFWGWG:22:SR}. This capability improves autonomous vehicles' decision-making in complex and crowded urban environments.

    \item \textit{Quantum Multi-Agent Cooperation}: Multi-agent \ac{QRL} frameworks are pivotal in environments like smart factories, where tasks such as autonomous robotic scheduling and resource optimization are key \cite{PKJK:24:APWCS}. These frameworks enhance inter-agent coordination and decision-making, achieving reduced computational overhead and improved task execution compared to classical multi-agent reinforcement learning.

    \item \textit{Autonomous Satellite Coordination}: \ac{QRL} has been applied to satellite-ground integrated systems, optimizing task allocation and dynamic resource management \cite{PKJK:24:APWCS}. Utilizing slimmable quantum neural networks, these systems adapt seamlessly to operational constraints and environmental changes, enhancing the performance of space-based autonomous networks. In addition, \cite{kim2025quantum} presents a QMARL model for coordinating multiple satellite systems, addressing the challenges associated with large-scale and high-dimensional tasks.

    \item \textit{Maze Optimization}: \ac{QRL} frameworks excel in navigation challenges such as the maze problem \cite{PBCM:22:QMI}. By leveraging quantum-enhanced exploration and decision-making, agents efficiently identify optimal paths through complex environments, outperforming classical reinforcement learning in terms of computational resource requirements and convergence speed.

    \item \textit{Collision Avoidance in Dense Environments}: Beyond self-driving cars, \ac{QRL} has been extended to manage dense traffic scenarios, modeling complex human interactions, and achieving real-time collision avoidance \cite{SFWGWG:22:SR}. By incorporating quantum-like Bayesian models, these systems address the unpredictability of human behavior, ensuring robust and safe navigation.

    \item \textit{Latent Space Optimization}: In hybrid quantum-classical reinforcement learning frameworks, \ac{QRL} has been applied to latent observation spaces for high-dimensional decision-making tasks, such as robotic and visual navigation \cite{NCBHKZ:24:arXiv}. These frameworks reduce computational overhead by compressing observations into latent representations, enabling efficient policy learning and improved scalability.
\end{itemize}

\paragraph{Optimization}
\ac{RL} is exceptionally effective in solving complex optimization tasks by enabling agents to learn optimal policies through iterative interactions. \ac{QRL} enhances this by integrating quantum computing for enhanced policy optimization. Notably, \ac{QRL} leverages methods such as Grover's search algorithm and parallel evaluation of state-action pairs, significantly reducing computational complexity and achieving superior results in decision-making tasks. Comparative studies show that \ac{QRL} not only matches but often surpasses classical deep \ac{RL} and quantum annealing approaches in challenging scenarios such as grid traversal \cite{NHP:23:QIP}. By utilizing gate-based quantum computing, \ac{QRL} demonstrates robust performance through Grover’s search for high-reward actions and parallel evaluation of state-action pairs, even under stochastic conditions. These strengths position \ac{QRL} as a pragmatic solution for optimization problems that are computationally prohibitive for classical methods. Below are key optimization tasks where \ac{QRL} has been effectively applied:

\begin{itemize}
    \item \textit{Combinatorial Optimization:} \ac{QRL} improves solution quality in problems like Weighted-MaxCut, Knapsack, and Unit Commitment by encoding the problems directly into Hamiltonians derived from their \ac{QUBO} forms \cite{KCRWL:24:arXiv}. The use of problem-specific quantum ansatz designs mitigates barren plateau issues, offering superior trainability and scalability compared to \ac{QAOA}, especially for generalizing across unseen problem instances.
    
    \item \textit{Two-Stage Decision Systems:} In renewable energy grids, \ac{QRL} can optimize day-ahead scheduling of thermal generators using Quantum Deep Q-Networks and handles real-time load adjustments with Quantum Soft Actor-Critic \cite{WZ:24:arXiv}. These quantum models balance cost and operational constraints under fluctuating renewable energy outputs, achieving robust task completion in a dynamic environment.
    
    \item \textit{Accelerator Beamline Control:} A hybrid actor-critic \ac{QRL} algorithm, integrating a quantum Boltzmann machine as the critic, was demonstrated to effectively optimize beam trajectories in CERN's proton and electron beamlines \cite{SCGKPLV:24:QST}. The approach utilizes quantum annealing for training, achieving faster convergence and adaptability in high-dimensional continuous action spaces.
    
    \item \textit{Stochastic Decision Problems:} \ac{QRL} can address the Frozen Lake problem, where stochastic transitions challenge classical RL models \cite{DMMML:22:arXiv}. By replacing neural networks with \acp{VQC} in Proximal Policy Optimization, \ac{QRL} achieves efficient representation and exploration of state-action spaces, requiring fewer parameters while maintaining robust learning.
    
    \item \textit{NFT-Based Intelligence Networking:} \ac{QRL} can optimize resource allocation in Non-Fungible Token (NFT)-based distributed intelligence systems for connected autonomous vehicles \cite{RXYHL:22:IEEE_M_NET}. Using quantum-enhanced policy optimization, vehicles dynamically decide retrieval modes and bandwidth allocation, minimizing delays while ensuring data integrity.
    
    \item \textit{Policy Optimization in Stochastic Tasks:} In grid traversal problems, comparative studies highlight \ac{QRL}'s advantage in sampling efficiency and convergence  \cite{NHP:23:QIP}. Gate-based \ac{QRL}, using Grover’s search, efficiently explores high-reward actions, while annealing-based \ac{QRL} achieves near-optimal policies through quantum-enhanced value estimation.
    
    \item \textit{Cloud-Based \ac{QRL}:} Quafu-RL \cite{BQG:23:arXiv}, implemented on a quantum cloud platform, trains agents using \acp{VQC} with hardware-efficient designs. For tasks like CartPole, Quafu-RL uses evolutionary architecture search to discover optimal circuit configurations, reducing gate count and improving training stability under noise.
    
    \item \textit{Resource Allocation in MEC:} \ac{QRL} can enhance joint task offloading and resource allocation in Mobile Edge Computing environments \cite{WGXYW:24:IEEE_J_MC}. Using a hybrid variational quantum-classical architecture, \ac{QRL} reduces the complexity of mixed discrete-continuous action spaces, achieving faster convergence and improved constraint compliance.
    
    \item \textit{Protein Folding:} \ac{QRL} can potentially solve the NP-complete protein folding problem by using \acp{VQC} to encode hydrophobic-pola lattice models  \cite{MI:23:SYNASC}. Through quantum policy updates, \ac{QRL} identifies near-optimal configurations while efficiently navigating the exponential search space.
    
    \item \textit{Multi-Agent UAV Networks:} \ac{QMARL} with quantum actor-critic networks can optimize large-scale UAV coordination tasks such as surveillance and mobile access  \cite{PK:23:APWCS}. By leveraging logarithmic action-space reduction through Projection Value Measure, \ac{QMARL} achieves robust convergence and scalability in multi-agent systems with high-dimensional state-action spaces.
\end{itemize}

\paragraph{Communication}
In communication, \ac{QRL} addresses critical challenges such as latency, resource allocation, and secure data transmission. It has demonstrated significant utility in enabling ultra-reliable, low-latency communication, dynamic task allocation, trajectory optimization, and privacy-preserving distributed learning. Applications span multiple domains, including \ac{UAV} networks, 6G systems, and energy trading. By efficiently modeling large, complex systems, \ac{QRL} provides scalable, adaptive solutions that surpass classical approaches in precision and computational efficiency in pushing the boundaries of 6G technologies and beyond \cite{WMDB:20:PRX_Quantum, CTRWTE:2024:Nat, PCPJCJK:24:IEEE_J_MC}.

\begin{itemize} 

\item \textit{Rediscovery and Optimization of Quantum Communication Protocols:} \ac{QRL} has been demonstrated to rediscover and enhance classical quantum communication protocols, such as teleportation and entanglement purification, particularly in non-ideal asymmetric conditions \cite{WMDB:20:PRX_Quantum}. It efficiently adapts to noise and stochastic environments, outperforming pre-designed classical protocols by dynamically optimizing fidelity and resource use.

\item \textit{Real-Time Adaptability in Distributed Networks:} \ac{QRL} enhances the integration of classical and quantum communication systems, enabling real-time decision-making in integrated networks such as \acp{SAGIN} \cite{CTRWTE:2024:Nat}. By leveraging quantum entanglement and teleportation, \ac{QRL} provides robust solutions for dynamic resource management and latency-sensitive applications. 

\item \textit{Spatio-Temporal Coordination for Metaverse Applications:} \ac{QRL} enables efficient spatio-temporal coordination in metaverse environments by integrating reinforcement learning with stabilized control \cite{PCPJCJK:24:IEEE_J_MC}. This ensures minimal latency and high-quality communication between virtual and physical systems.

\item \textit{Blockchain-Integrated \ac{QRL} for Secure Energy Trading:} In decentralized systems like e-mobility energy trading, \ac{QRL} combined with blockchain, can optimize resource allocation and secure data exchange \cite{KDKK:23:IEEE_J_VT}. By leveraging smart contracts and dynamic pricing mechanisms, \ac{QRL} ensures low latency and transparent energy allocation, enhancing trustworthiness and efficiency.

\item \textit{Improving \ac{UAV} Communication and Coordination:} \ac{QRL}-based frameworks improve \ac{UAV} trajectory optimization by enhancing sampling efficiency and reducing computational overhead \cite{LAD:22:IEEE_J_WCOM}. Through Grover-inspired experience replay and dynamic action space adjustments, \ac{UAV} systems achieve better synchronization and stability in trajectory planning and communication. 

\item \textit{Integrated Sensing and Communication (ISAC):} In ISAC systems, \ac{QRL} enhances tasks such as direction-of-arrival estimation and task offloading by optimizing the trade-offs between sensing and communication \cite{PSKCLDD:24:IEEE_J_JSAC}. In particular, the quantum actor-critic approach results in lower latency and higher fidelity in real-time scenarios like surveillance and defense systems.

\item \textit{Optimizing Digital Twin Deployment in 6G Networks:} Multi-agent \ac{QRL} frameworks address the challenge of deploying digital twins in edge computing environments, reducing latency while meeting computational constraints \cite{TUSPSK:24:INIS}. By leveraging amplitude encoding, \ac{QRL} scales efficiently, ensuring dynamic updates in complex 6G networks. 

\item \textit{Joint Optimization of \ac{UAV} Trajectory and Resource Allocation:} \ac{QRL} has been applied to jointly optimize \ac{UAV} trajectories and resource allocation in high-mobility environments \cite{NSO:23:IEEE_J_IOT}. This approach reduces energy consumption and ensures stable communication by leveraging quantum layers in neural networks, achieving improved latency and scalability.

\end{itemize}

\paragraph{Finance}

Finance is an inherently complex and constantly evolving domain, influenced by volatile markets and many unpredictable factors. This is especially evident in market making, portfolio management, and order execution, where conditions change within seconds and require continuous adaptation and fast decision-making \cite{bai2025review}. Traditional machine learning models often fail to cope with such rapid and dynamic environments. RL has been increasingly applied in finance, allowing agents to learn through interaction, adapt to changing market conditions, and optimize decisions over time. Because quantum computing is anticipated to bring its earliest practical benefits to finance \cite{herman2022survey}, researchers have begun exploring QRL as a way to further enhance adaptability and decision-making in complex financial settings. Several recent studies have explored the application of QRL across different financial domains, including:  

\begin{itemize}
    \item \textit{Deep Hedging:} In \cite{raj2023quantum}, QRL methods were developed for Deep Hedging. In particular, quantum neural networks with orthogonal and compound layers were used to represent policy and value functions. In addition, a distributional actor-critic algorithm was proposed, which leverages the large distributions that come with quantum states.
    \item \textit{Algorithmic Trading:} The integration of Quantum Long Short-Term Memory (QLSTM) and QRL for algorithmic trading was proposed in \cite{chen2025quantum}. The authors combined QLSTM for short-term trend forecasting with Quantum Asynchronous Advantage Actor–Critic (QA3C) for decision-making, creating a hybrid model capable of learning both predictive patterns and trading strategies. The QLSTM acted as a feature extractor of market trends, which were then used as state inputs to the QA3C agent.
    \item \textit{Optimizing Fintech Trading Decisions:} Classical LSTM was integrated with QA3C for S\&P 500 trading in \cite{liu2025quantum}. The LSTM model is used to generate one-week-ahead forecasts of macroeconomic and price features, which are then fed as additional predictive inputs to QA3C agents.
    
\end{itemize}

\paragraph{Quantum Architecture Search}

While classical reinforcement learning has been successfully applied to optimize quantum architectures, as discussed in Section~\ref{QAS}, recent work explored QRL, where quantum agents interact with quantum environments to optimize circuit design and control. In \cite{chen2023quantum2}, the quantum agent operates in a quantum environment where its actions correspond to selecting quantum gates and operations to build a circuit. After constructing a candidate circuit, the agent receives a reward based on performance metrics, guiding it to favor better architectures. However, research in this area remains limited, and most existing work still employ classical RL for QAS, or use QAS methods to improve QRL agents, such as in \cite{chen2024differentiable,sun2023differentiable}. 

\section{Future Directions and Open Problems}\label{sec:future}
\Ac{QML} has recently garnered significant attention due to its ability to address well-known challenges in classical machine learning, such as scalability and computational bottlenecks. The availability of robust \acp{SDK}, including Qiskit, TensorFlow Quantum, PennyLane, and curated datasets and benchmarks, has made \ac{QML} relatively accessible to a wider audience. However, the entry barrier for \ac{QML} remains moderate to high, demanding a solid understanding of quantum mechanics and classical machine learning frameworks. \ac{QRL} faces even higher entry barriers as it further requires advanced expertise in reinforcement learning and optimization techniques. Additionally, \ac{QRL} remains a niche field, with adoption hindered by several challenges:

\begin{enumerate}
    \item \textbf{High Complexity}: The multidisciplinary nature of \ac{QRL} demands an advanced understanding of quantum mechanics, reinforcement learning algorithms, and optimization methodologies. This complexity limits its accessibility to researchers and practitioners.

    \item \textbf{Limited Resources}: Unlike \ac{QML}, \ac{QRL} suffers from a lack of specialized \acp{SDK}, curated datasets, and standardized benchmarks. This scarcity inhibits experimentation and hinders its growth within the research community.

    \item \textbf{Hardware Constraints}: Practical implementations of \ac{QRL} algorithms often require sophisticated quantum hardware. Current technological limitations, such as qubit coherence and error rates, pose significant challenges to executing \ac{QRL} at scale.

    \item \textbf{Niche Status}: \ac{QRL} has not yet achieved widespread adoption due to the above, making it less appealing than other machine learning paradigms such as neural networks and support vector machines.
\end{enumerate}

While \ac{QRL} faces significant challenges due to its complexity and lack of resources, it holds the potential for addressing unique problems where quantum advantages can be leveraged. As quantum hardware continues to improve and more resources become available, \ac{QRL} is poised to unlock new possibilities in machine learning and beyond. As \ac{QRL} continues to gain traction, some pressing challenges remain that slow its wide adoption. In the following, we highlight the main limitations and open problems in \ac{QRL}

\begin{table*}[ht]
\centering
\caption{Neural Network Architectures}
\begin{tabularx}{\textwidth}{l X X}
\toprule
\textbf{Architecture} & \textbf{Description} & \textbf{Mathematical Representation} \\
\midrule
Multilayer Perceptron (MLP) [1958] & 
Feedforward neural network using weighted sums and non-linear activation functions. &
$h^{(l)} = \sigma(W^{(l)} h^{(l-1)} + b^{(l)})$ \\
Convolutional Neural Network (CNN) [1980] & 
Neural network leveraging convolutional layers for spatially localized patterns. &
$h^{(l)}_{i,j} = \sigma\left( \sum_{m,n} W^{(l)}_{m,n} \cdot h^{(l-1)}_{i+m,j+n} + b^{(l)} \right)$ \\
Tensor Networks [1992] & 
Efficient representation of high-dimensional tensors through decompositions. &
$|\psi\rangle = \sum_{i_1,\ldots,i_N} \text{Tr}(A^{[1]}_{i_1}  \cdots A^{[N]}_{i_N}) |i_1 \ldots i_N\rangle$ \\
Long Short-Term Memory (LSTM) [1997] & 
Recurrent neural network variant with gating mechanisms for handling sequential data. &
$\begin{aligned}
f_t &= \sigma(W_f x_t + U_f h_{t-1} + b_f) \\
i_t &= \sigma(W_i x_t + U_i h_{t-1} + b_i) \\
o_t &= \sigma(W_o x_t + U_o h_{t-1} + b_o) \\
\tilde{c}_t &= \tanh(W_c x_t + U_c h_{t-1} + b_c) \\
c_t &= f_t \odot c_{t-1} + i_t \odot \tilde{c}_t \\
h_t &= o_t \odot \tanh(c_t)
\end{aligned}$ \\
Transformer [2017] & 
Self-attention mechanism for sequence-to-sequence tasks. &
$\text{Attention}(Q, K, V) = \text{softmax}\left(\frac{QK^T}{\sqrt{d_k}}\right)V$ \\
Continuous-Variable Quantum Neural Networks [2018] & 
Quantum states processed using continuous-variable parameterized quantum gates. &
$\begin{aligned}
\ket{\psi_{\text{out}}} &= \prod_{i} D(\alpha_i) S(r_i) R(\phi_i) \ket{\psi_{\text{in}}}, \\
D(\alpha) &= e^{\alpha \hat{a}^\dagger - \alpha^* \hat{a}}, \\
S(r) &= e^{\frac{r}{2} (\hat{a}^2 - \hat{a}^{\dagger 2})}, \\
R(\phi) &= e^{-i \phi \hat{n}}
\end{aligned}$ \\
Convolutional Differentiable Logic Gate Networks [2020] & 
Differentiable logic gates (e.g., NAND, OR) combined with learnable weights for computation. &
$h_j^{(l)} = \sigma\left(\sum_{i} w_{ij}^{(l)} \cdot g_{ij}^{(l)}(h_i^{(l-1)})\right)$ \\
Kolmogorov-Arnold Networks (KAN) [2024] & 
Learnable univariate functions applied to connections between neurons for dynamic adaptability. &
$h_j^{(l)} = \sum_{i} f_{ij}^{(l)}(h_i^{(l-1)})$ \\
\bottomrule
\end{tabularx}
\label{tab:math_representations}
\end{table*}

\label{sec:future}
\subsection{\ac{QRL} Architectures}
The architecture of \ac{QRL} is crucial as it determines its capacity to learn, generalize, and perform effectively across various tasks. Key architectural choices, such as the selection of parameters, activation functions, and computational gates, significantly influence the network's performance and suitability for specific applications. In the following, we discuss recent design developments in neural network architectures that can be used within a \ac{QRL} setting, drawing inspiration from classical learning paradigms as shown in Table~\ref{tab:math_representations}.

\paragraph{\ac{KAN}} Inspired by the Kolmogorov-Arnold representation theorem,  which states that any multivariate continuous function can be represented as a finite composition of continuous univariate functions and addition \cite{LWVRHSH:24:arXiv}. In \acp{KAN}, each connection between neurons is associated with a learnable univariate function, often parameterized as a spline, allowing dynamic adaptation to complex data patterns. The application of \ac{KAN} in quantum computing has been effectively demonstrated in \ac{QML} frameworks, where it enhances tasks like quantum state preparation and designing \acp{VQC}. For instance, \cite{KSS:24:EPJ_QT} shows that \ac{KAN} enables the design of compact \acp{VQC} with fewer two-qubit gates and reduced depth, addressing major limitations of current \ac{NISQ} devices, such as noise sensitivity and short coherence times. Additionally, \cite{IHKPR:24:arXiv} highlights that \ac{KAN}'s learnable activation functions and efficient parameterization outperform traditional \acp{MLP}, offering robustness and scalability to larger quantum systems. Future research should extend \ac{KAN}-based \ac{QRL} to multi-task hybrid quantum-classical learning, improve the interpretability of learned functions, and reduce execution times using specialized hardware accelerators, broadening its impact on practical quantum computing applications.

\paragraph{Convolutional Differentiable Logic Gate Networks}
\Acp{CDLGN} are a novel machine learning architecture that integrates the efficiency of logic gate operations with the representational power of convolutional neural networks. Using differentiable relaxations of logic gates such as NAND, OR, and XOR, \acp{CDLGN} enables gradient-based optimization, facilitating direct learning of logic gate configurations for specific tasks. This approach allows for the construction of models that perform inference using only logic gate operations, which are inherently faster and more hardware-efficient than traditional neural network computations. In a recent work \cite{PKBWE:24:arXiv}, researchers demonstrated the advantages of \acp{CDLGN} by achieving an accuracy of $86.29\%$ on the CIFAR-$10 $dataset using only $61$ million logic gates. This performance surpasses previous state-of-the-art models while being $29$ times smaller in terms of gate count, highlighting the efficiency and scalability of \acp{CDLGN}. This provides an opportunity to explore whether integrating \acp{CDLGN} can enhance quantum-inspired \ac{RL} by enabling rapid decision-making and policy evaluations with reduced computational overhead. By leveraging their efficient inference capabilities, \acp{CDLGN} has the potential to streamline intensive operations. Furthermore, their inherent interpretability could offer deeper insights into decision-making processes, leading to improved performance and transparency of \ac{RL} agents. Incorporating \acp{CDLGN} into \ac{RL} frameworks could drive significant advancements in both efficiency and understanding.

\paragraph{Continuous-variable quantum neural networks} \acp{CV-QNN} are a class of quantum neural networks operating within the continuous-variable framework of quantum computing. Unlike traditional qubit-based systems, they encode information in continuous degrees of freedom, such as the amplitudes and phases of electromagnetic fields, making them well-suited for tasks involving continuous data. \acp{CV-QNN} can implement nonlinear activation functions through non-Gaussian operations, enabling the construction of universal quantum computation models \cite{KBASQL:19:PRR,BSYS:23:PRA}.  Despite their implementation complexities, such as precise control over continuous quantum states and maintaining coherence, \acp{CV-QNN} offer significant advantages. They naturally process continuous data, facilitate encoding for quantum algorithms, and leverage high-dimensional quantum entanglement for powerful computational models.  Frameworks such as Strawberry Fields \cite{KIQBAW:19:Quantum} and Piquasso \cite{KRRKPJNVMEM:24:arXiv} are instrumental in designing novel \ac{QRL} architectures. Strawberry Fields provides tools for constructing, simulating, and optimizing continuous-variable quantum circuits, while Piquasso offers a flexible platform for modeling and simulating continuous-variable quantum systems. By utilizing these frameworks, researchers can explore \ac{QRL} architectures that harness the unique capabilities of \acp{CV-QNN}, driving advancements in efficiency and interpretability.

\paragraph{Tensor Networks}
Tensor networks are mathematical structures that decompose high-dimensional tensors into interconnected lower-dimensional tensors, enabling efficient representation and computation of complex data. They are particularly effective in modeling quantum many-body systems by capturing intricate correlations and entanglements \cite{W:92:PRL}. In the context of \ac{QRL}, tensor networks present a promising solution to the scalability challenges inherent in \ac{QRL} algorithms. Scaling \ac{QRL} is computationally intensive due to the exponential growth of quantum state spaces and execution times on quantum hardware, as highlighted in recent studies \cite{ML:24:arXiv}. By leveraging tensor networks, these large state spaces can be approximated and managed more efficiently, facilitating the design of scalable and effective \ac{QRL} architectures \cite{RKR:23:PRSA}. In \cite{chen2022variational}, this was further demonstrated through a hybrid tensor network variational quantum circuit architecture that combines matrix product states with variational quantum circuits for reinforcement learning tasks. Additionally, the integration of RL and tensor networks has demonstrated their potential to enhance scalability and performance in quantum learning models. Recent work explored combining RL with tensor networks to address dynamical large deviations, showcasing the versatility of tensor networks in improving computational efficiency \cite{GRG:24:PRL}. Tensor networks offer a pathway to develop practical and efficient \ac{QRL} frameworks, addressing the critical challenge of execution time and resource consumption.

\paragraph{Quantum-Train}

Quantum-Train (QT) integrates quantum computing with classical machine learning algorithms by using Quantum Neural Network (QNN) during training to generate or optimize the parameters of a classical neural network (NN) \cite{liu2405quantum}. This framework addresses key issues in QML, such as limited access to quantum hardware and information loss in data encoding. Moreover, QT significantly reduces the number of parameters required to train classical NNs. This could have a huge advantage in the context of QRL, where model efficiency and scalability are critical. Similar concepts have been explored in the literature, where a QNN is used only during training to generate the parameters of a classical policy network \cite{liu2024qtrl}. This work has been extended to QT-Based Distributed Multi-Agent Reinforcement Learning, where multiple QPUs were used for parallel training and parameter synchronization \cite{chen2025quantum2}. These results highlight a promising direction for future \ac{QRL} research, where quantum parameter generation is leveraged  to build scalable, efficient, and hardware-feasible reinforcement learning systems.

\paragraph{Adaptive Non-Local Observables}

A recent direction in \ac{QRL} architecture design focuses on enhancing the measurement layer of variational quantum circuits rather than deepening the circuit itself. The authors in \cite{lin2025quantum} introduced Adaptive Non-Local Observables (ANO) for quantum reinforcement learning to address the limitations of local measurements. ANO optimizes both the circuit parameters and multi-qubit measurements. The proposed architecture significantly expands the function space of quantum agents without increasing the depth. When incorporated into DQN and A3C frameworks, ANO-VQC agents achieved faster convergence and higher cumulative rewards than traditional VQCs. Future research could investigate the integration of adaptive observables with other architectural paradigms.

\subsection{LLM and \ac{QRL}}
\Acp{LLM} have become instrumental in code generation, significantly enhancing developer productivity and reducing the learning curve for new developers \cite{DMNKDJ:23:JSS, PKCD:2023:arXiv, I:2022:ICSE_CP}. While general-purpose models such as StarCoder, Code Llama, and DeepSeek Coder have demonstrated strong performance across conventional programming benchmarks, they encounter substantial limitations in specialized quantum domains, where intricate domain-specific knowledge is essential \cite{LLACLTPLYW:2024:arXiv, RGGSGTALS:2023:arXiv, GZYXDZCBWL:2024:arXiv}. Popular quantum \acp{SDK}—such as Qiskit, Cirq, PennyLane, and OpenQASM—are deeply rooted in quantum mechanics, making them indispensable for navigating the complexities of quantum circuits and supporting the development of advanced quantum algorithms \cite{JTKWLGMBNC:2024:arXiv, BISGAAAANAA:2018:arXiv, CJAABHRSG:22:ACM_TQC}. Beyond general-purpose SDKs, domain-specific tools for applications such as quantum sensing (e.g., OQuPy\cite{FFFBBEGKKL:24:JCP}, quantum control (e.g., QuTiP\cite{JNN:12:CPC}), and quantum communication (e.g., NetSquid\cite{CKDMNOPRRS:21:Comm_Phys}) are pivotal in addressing unique challenges across these fields. The emergence of quantum-specific code assistants reflects the growing demand for tools that can bridge the gap between general-purpose \acp{LLM} and domain-specific requirements. For instance, Qiskit has established itself as a versatile \ac{SDK}, supporting programming across multiple abstraction levels, from high-level algorithmic design to low-level quantum gate manipulation. Its modular structure enables circuit optimization and hardware retargeting, allowing compatibility across diverse quantum architectures. Enhancing accessibility, the Qiskit Code Assistant provides tailored code snippets for users with minimal quantum programming experience \cite{DBVFKFPC:24:arXiv}. Similarly, KetGPT augments training datasets with synthetic quantum circuits that mimic real-world algorithms, enriching \acp{LLM} with quantum-specific capabilities and improving their precision in generating quantum instructions \cite{ABSF:2024:ICCS}. To evaluate \acp{LLM} tailored for quantum programming, benchmarks such as Qiskit HumanEval and QASMBench have been developed. Qiskit HumanEval includes over 100 tasks, covering quantum circuit generation, state preparation, and algorithmic implementations, setting a high standard for functional accuracy and executable code generation \cite{VHGBCDFPC:2024:arXiv}. Meanwhile, QASMBench targets low-level OpenQASM benchmarks, focusing on metrics like gate fidelity, circuit depth, and noise resilience across platforms such as IBM-Q and Rigetti \cite{LSKA:2023:ACM_TQC}. Broader frameworks, like MQT Bench, span abstraction layers from algorithm design to hardware-specific deployment, measuring performance metrics such as two-qubit gate count and circuit depth across multiple quantum processors \cite{QBW:23:Quantum}. Additionally, benchmarking efforts like VHDL-Eval and L2CEval extend evaluation into specialized domains such as hardware description languages and multi-domain code generation tasks \cite{NYZRFYSYXL:2024:TACL, VSAMNBD:2024:arXiv}.

Drawing inspiration from these advancements, the development of \ac{QRL} agents emerges as a natural progression. \Acp{QRL} aim to harness the principles of quantum mechanics and RL to create agents capable of navigating quantum environments. However, designing effective \ac{QRL} agents necessitates integrating tools from both quantum computing and reinforcement learning. These agents must support the efficient formulation of quantum states, application of quantum gates, and optimization of quantum circuits while interacting with quantum environments to receive feedback and adjust strategies. Benchmarks similar to Qiskit HumanEval or QASMBench can be adapted for \ac{QRL} tasks to evaluate agents' performance in quantum state preparation, gate optimization, and reinforcement learning-specific objectives. By building on these foundations, \ac{QRL} agents can unlock new frontiers in quantum machine learning, providing scalable and efficient solutions for quantum algorithms.

\subsection{Quantum-centric supercomputing}
Quantum-centric supercomputing refers to a hybrid computational paradigm in which quantum processors are seamlessly integrated with classical high-performance computing systems, leveraging quantum capabilities to accelerate specialized tasks within a unified architecture. The potential of \ac{QRL} and quantum-inspired \ac{RL} to enable quantum-centric supercomputing lies in their ability to bridge classical and quantum paradigms, optimizing both hardware utilization and algorithmic design\cite{MSAJBCCEFMKLMOPSSVTW:24:arXiv}. A representative example of such systems is presented in \cite{park2025s}, where distributed quantum convolutional networks operate on separate quantum processors and are classically aggregated within a Double Deep Q-Network framework. This design demonstrates scalable quantum workload distribution and efficient handling of high-dimensional data.

These frameworks can play a pivotal role in realizing scalable and efficient quantum systems by addressing key challenges:

\begin{itemize}
    \item \textit{Optimization of Hybrid Systems}: 
    \ac{QRL} integrates classical reinforcement learning with quantum operations, facilitating the dynamic optimization of hybrid quantum-classical workloads. This improves resource allocation, reduces bottlenecks, and accelerates fault-tolerant quantum computation tasks.

    \item \textit{Quantum Workload Distribution}: 
    Quantum-inspired reinforcement learning can effectively manage workload distribution across quantum and classical coprocessors, including \acp{QPU} and GPUs. Adaptive circuit knitting methods further enhance this capability, making \ac{QRL} instrumental in coordinating computations in quantum-classical systems.

    \item \textit{Enhanced Training and Calibration}: 
    \ac{QRL} agents automate the recalibration of quantum devices, minimizing coherence loss and mitigating error accumulation. This capability is critical for maintaining the performance of large-scale quantum systems.

    \item \textit{Algorithmic Advancement}: 
    Quantum-inspired reinforcement learning fosters the development of heuristic algorithms optimized for \ac{NISQ} devices and beyond. These algorithms address the inherent noise and limited qubit count of current quantum systems while preparing for the transition to utility-scale quantum supercomputers.

    \item \textit{Scalability and Fault Tolerance}: 
    \ac{QRL} aids in designing strategies for fault-tolerant logical qubit operations and efficient use of quantum error correction codes, significantly reducing the overhead of scaling up to millions of physical qubits required for utility-scale quantum supercomputing.
\end{itemize}


\section{Conclusion}
\label{sec:conclusion}
In this survey, we highlight the potential of \ac{QRL} in advancing quantum computing and its integration with classical systems. By leveraging core principles of quantum mechanics—such as superposition and entanglement—\ac{QRL} frameworks enable more effective exploration, policy learning, and optimization for complex decision-making tasks. The use of variational quantum circuits in these frameworks addresses challenges such as noise and limited coherence times in \ac{NISQ} devices, positioning \ac{QRL} as a viable approach to achieving near-term quantum advantage. Recent developments demonstrate the versatility of \ac{QRL} across multiple domains, including quantum architecture search, quantum sensing, optimization problems, and autonomous systems in classical contexts. Key innovations, such as learnable activation functions in \ac{KAN}-based architectures, adaptive circuit knitting techniques, and efficient hybrid workload management, highlight its potential to enhance scalability and computational efficiency across both quantum-specific and interdisciplinary applications.

However, \ac{QRL} still faces several challenges. These include algorithmic complexity, hardware limitations, and the lack of standardized tools and benchmarks. Overcoming these obstacles will require the development of accessible software frameworks, the integration of domain-specific tools, and the creation of \ac{QRL} assistants to lower the barriers for researchers and practitioners entering the field. Looking ahead, \ac{QRL} may play a pivotal role in enabling quantum-centric supercomputing, a hybrid approach that integrates quantum and classical resources. By advancing hybrid system optimization, automating quantum device calibration, and developing scalable fault-tolerant systems, \ac{QRL} can facilitate progress toward utility-scale quantum computing. As quantum hardware and algorithmic frameworks mature, \ac{QRL} has the potential to enable significant advancements across scientific and industrial domains.

\balance

\bibliographystyle{IEEEtran}
\bibliography{
IEEEabrv, 
./LatexInclusion/IEEE-TPAMI-QRL
}

\end{document}